# Phonon-assisted exciton / trion conversion efficiency in transition metal Dichalcogenides

Sabrine Ayari[1], Sihem Jaziri[1,2], Robson Ferreira[3] and Gerald Bastard[3]

[1]Faculté des Sciences de Bizerte, Laboratoire de Physique des Matériaux
Structure et Propriétés , Université de Carthage,7021 Jarzouna, Tunisia

[2]Faculté des Sciences de Tunis, Laboratoire de Physique de la Matiére Condensée, Département
de Physique, Université Tunis el Manar, Campus Universitaire 2092 Tunis, Tunisia

[3]Laboratoire de Physique de l' école normale supérieure, ENS, Université PSL, CNRS,
Sorbonne Université, Université Paris-Diderot, Sorbonne Paris Cité, Paris, France

(Dated: May 10, 2020)

**Abstract**

Photoluminescence spectra, shows that monolayer Transition-metal dichalcogenides (ML-TMDCs), possess charged exciton binding energies, conspicuously similar to the energy of optical phonons. This enigmatic coincidence has offered opportunities to investigate many-body interactions between trion $(X_-)$, exciton $(X)$ and phonon and led to efficient excitonic anti-Stokes processes with the potential for laser refrigeration and energy harvesting. In this study, we show that in WSe$_2$ materials, the trion binding energy matches two phonon modes, the out-of-plane $A'_1$ and the in-plane $E'$ mode. In this respect, using the Fermi golden rule together with the effective mass approximation, we investigate the rate of the population transfers between $X$ and $X_-$, mediated by a single phonon. We demonstrate that, while the absolute importance of the two phonon modes on the up-conversion process strongly depend on the experimental conditions such as the temperature and the dielectric environment (substrate), both modes lead to an up-conversion process on time scales in the range of few picoseconds to sub-nanosecond, consistent with recents experimental findings. The conjugate process is also investigated in our study, as a function of temperature and electron density $N_e$. We prove that exciton to trion down-conversion process is very unlikely at low electron density $N_e < 10^{10}\ cm^{-2}$ and high temperature $T > 50\ K$ while it increases dramatically to reach few picoseconds time scale at low temperature and for electron density $N_e > 10^{10}\ cm^{-2}$. Finally, our results show that conversion process occurs more rapidly in exemplary monolayer molybdenum-based dichalcogenides (MoSe$_2$ and MoTe$_2$) than tungsten dichalcogenides .

## I. INTRODUCTION AND MOTIVATION

Transition-metal dichalcogenides (TMDCs) (MX2, M=Mo,S; X=Se,S,Te), as atomically thin two- dimensional materials have had a tremendous impact on semiconductor physics in the last years. Due to their direct bandgap located at the $\pm$K points in the Brillouin zone, a large interband dipole moment, spin-valley coupled physics, atomically thin layers emit strong photoluminescence and represent a semiconducting supplement to the two-dimensional and zero-gap material grapheme [1-7]. The attractive Coulomb interaction between the conduction band electron and the valence band hole in TMDCs can bound them into a hydrogen-like state, known as exciton *X*, which is an elementary excitation that plays a key role in optoelectronic phenomena [8-15]. In TMDCs materials, owing to the dielectric screening and spatial quantum confinement, their optical transitions and light-matter interaction are governed by robust exciton feature with binding energies of several hundred meV [8-14]. This strong binding energy, which is more than one order of magnitude larger than conventional semiconductors such as GaAs, leads to large optical transition dipoles of excitons [8-14, 16, 17]. Additionally, when the TMDCs sample is negatively doped, the bound electron-hole pair can also capture an extra electron to form a negatively charged exciton, also known as trion, denoted by $X_-$, with binding energy in the order of tenth meV allowing the observation of their photoluminescence up to the room temperature [2, 9, 11, 15, 17, 18].

Recent experiments have demonstrated the existence of efficient up-conversion processes leading to anti-Stokes photoluminescence (ASPL) in semiconducting TMDCs monolayers [17, 19-26]. ASPL has great potential for applications in the multi-color displays [27], dynamical imaging microscopy [28], unconventional lasers [29], bio-imaging agents [30] and solid-state laser cooling devices [31]. The up-conversion photoemission has been observed in rare earth atoms [32], in glasses [33], carbon nanotubes [34] and for different semiconductor structures such as InP/InAs heterojunctions, CdTe quantum wells, InAs quantum dots and atomically thin semiconductors at cryogenic temperatures [35-39]. Actually, TMDCs are particularly promising for room temperature up-conversion due to their very strong photon-exciton and phonon-exciton interactions [17, 19-21, 24, 26].

The origin of the up-conversion signal is in itself a crucial problem, as the origin of excess energy needs to be identified. Indeed, the discovery of up-conversion process in semiconducting monolayer TMDCs has triggered a variety of experimental and theoretical studies seeking to understand its microscopic origin. Surprisingly, it has been shown that the microscopic origin of the anti-Stokes emission process in TMDCs strongly depend on the experimental condition, such as temperature, excitation power, doping density and type of the sample (Mo/W) [17, 19-21, 23-26, 40]. For T $\geq$ 30 K and low excitation density, Jones *et al.*[17] have demonstrated the up-conversion process from a negatively charged exciton ($X_-$, resonantly excited at 1720 meV) to a neutral exciton (*X* luminescence at 1750 meV) in monolayer $WSe_2$, producing spontaneous anti-Stokes emission with an energy gain of 30 meV. They ascribed this spontaneous excitonic anti-Stokes to doubly resonant Raman scattering mediated by the $A'_1$ optical phonon. In the same context, J. Jadczak *et al.*[24] have reported a room temperature prominent up-conversion photoluminescence process in a monolayer semiconductor $WS_2$, with energy gain up to 150 meV. Similarly as in ref.[17], they attributed the up-conversion to the excitonic anti-Stokes Raman scattering process. However, in their experiments the up-conversion is related to multi-phonon assisted transition since the incident photon energy is well below (in the energy tail) of the trion. In these papers [17, 24], as all the data of total up-conversion intensities follow sublinear or linear dependencies with the exciting power, the authors rule out the possibility of nonlinear optical processes such as two-photon excitation and exciton Auger scattering. Other authors, like M. Manca *et al.* [21] and B. Han *et al.* [19] have demonstrated that for TMDCs monolayers encapsulated in hexagonal boron nitride (h-BN), at low temperature and for moderate excitation powers, the resonant excitation of the lowest-energy exciton $A_{1s}$ results in ASPL emissions from both the B-exciton, (up-shifted of 430 meV for WSe$_2$ or 240 meV for MoSe$_2$) and the excited A$_{2s}$ exciton (up-shifted of 130 meV for WSe2 or 148 meV for MoSe2 ). The authors interpret the observed generation of highly excited excitons within a double scenario involving resonant two-photon absorption and Auger-type shake up processes since the up-conversion intensities scale quadratically with the photo-excitation intensity [19, 21]. They discard any phonon-assisted contribution to the ASPL signal, since the up-shifts are much higher than the TMDCs phonons energies and, at low temperature (4 K) the thermal phonon energies are less than 0.3 meV. In this work, we theoretically evaluate the efficiency of the trion-to-exciton up-conversion in terms of a sequential process assisted by the absorption of an optical phonon. To this aim, we shall use the Fermi golden rule to calculate within the effective mass approximation the rate of the population transfers between X and $X_-$ mediated by resonant phonons. We shall consider in detail the dependencies of the trion to exciton rate on several parameters of interest (such as the temperature of the crystal lattice and the dielectric environment) for both families of TMDCs material : tungsten and molybdenum TMDCs ML. In addition, the exciton $X$ to trion $X_-$ down conversion process assisted by one phonon emission is also investigated in our work. In this context, we provide here an in depth study of the population transfer efficiency from exciton to trion as a function of temperature, electron density and TMDCs materials.

## II. PHONON ASSISTED TRION TO EXCITON UP CONVERSION PROCESS

TMDCs usually present a significant residual n-doping. As a consequence, both the absorption type (reflectivity contrast) and the PL emission spectra display a clear low energy (as compared to the one of the intrinsic exciton (X) transition) peak associated with the trion (negative charged exciton $X_-$) recombination. It has been shown that the PL emission intensity of the latter strongly depends on the temperature and two dimensional carrier gas concentration [15, 17, 24, 26, 40-44]. In this work, we will address the usual situation where the TMDCs ML samples are micromechanically cleaved from a bulk of TMDCs crystal and deposited on a SiO$_2$ layer on top of a Si substrate. Additionally, we shall assume low doping regime ($N_e \leq 10^{12}\ cm^{-2}$) and weak optical excitation conditions. More precisely, we will focus on the exchange of populations between the trion and the dissociated exciton (exciton plus one free electron), due to the couplings of the electrons and hole to optical phonons. The model is compared to the experimental data on ML-TMDCs available in the literature.

### A. Theoretical formulation of the up-conversion process

In this section, we provide a theoretical model to calculate the up-conversion rate in TMDCs ML. In this process, one quantum of lattice oscillations (optical phonon) is absorbed. Here, our theory and interpretation are based on the recent work of Jones *et al*. [17] and K.Hao *et al*.[26] in which they demonstrated an efficient luminescence up-conversion process from a trion to an exciton resonance in monolayer WSe$_2$ and MoSe$_2$, respectively, producing spontaneous anti-Stokes emission with an energy gain of 30 meV. As depicted in Fig.1(a), the absorption of a **Q**-wave vector optical phonon from the bath, promotes a trion $X_-$ (bound electron-exciton) with center of mass wave vector $K_{X_-}^i$ into a final state composed of an unbound (free) electron with wave vector $\boldsymbol{q_e}$ and one neutral exciton $X$ with center of mass wave vector $\boldsymbol{K_X}$. Based on the Fermi golden rule, the trion-to exciton up-conversion rate $W_{\lambda,X_-\to X}^{UC}(K_{X_-}^i)$ summing over all the final states $\boldsymbol{q_e}$, $\boldsymbol{K_X}$, and over all available initial phonon wavevectors **Q** is given as follow :

$$W^{UC}_{\lambda,X_-\to X}\left(K^i_{X_-}\right) = \frac{2\pi}{\hbar}\Sigma_{Q,q_e,K_X}\left|F^{UC}_\lambda(Q,q_e,K_X K^i_{X_-})\right|^2 \delta\left(E_F(K_X,q_e,Q) - E_I(K^i_{X_-},Q)\right) \qquad (1)$$

where, $\lambda$ is the optical phonon mode, the $\delta$ function ensures the conservation of energy during the scattering process. $E_I(K^i_{X_-},Q)$ and $E_F(K_X,q_e,Q)$ re the initial and final energies for the trion and the electron-exciton states, respectively. The coupling strength is given by the matrix element for single optical-phonon $jth$ particle scattering process:

$$F^{UC}_\lambda(Q,q_e,K_X,K^i_{X_-}) = \langle\psi_F|H^\lambda_{X_--ph}|\psi_I\rangle \quad (2)$$

In the very first step, in order to calculate phonon-assisted trion to-exciton up-conversion rate, there are three physical quantities that should be determined i) The energy separation between exciton and trion state (trion binding energy) with a view to identify the optical phonon involved in the scattering process (in Section A.1). ii) The trion phonon interaction potential (in Section A.2). iii) the initial and final states involved in the up-conversion process (in Section A.3).

**1. Energy separation between neutral exciton and trion states**

The motivation for our studies of the exciton complex-phonon interactions was based on the comparable values of the energy separation of the $X$ and $X_-$ lines in PL spectra of the common monolayer TMDCs (MoSe$_2$, WSe$_2$, WS$_2$, MoS$_2$, MoTe$_2$) with the energies of optical phonons. Hence, before investigating the trion-exciton up-conversion rate, we will first calculate the trion binding energy $E_{BX_-} = E_X - E_{X_-}$, where $E_X$ and $E_{X_-}$ are the energy of the exciton and trion, with vanishing centre-of-mass wave vectors respectively. To estimate the binding energies and the oscillator strength of the exciton and trion states in monolayer TMDCs, we present in Appendix.A our theoretical framework based on the 2D effective-mass approximation. The values of the exciton and trion energies as well as the trion binding energies for the five most commonly investigated TMDCs semiconductors MoS$_2$, MoSe$_2$, WS$_2$, and WSe$_2$, MoTe$_2$ (deposited on the SiO$_2$/Si substrate with high-frequency dielectric constants $\varepsilon_b = 2.1$ and exposed to the air with $\varepsilon_t = 1$, see discussion in Appendix B) are summarized in Table I. For comparison, we report in the same table, the phonon energies of the longitudinal optical (LO) $E'$ phonon and out-of-plane transverse $A'_1$ (HP) optical phonons extracted from the literature for all MLs (see the discussion in ref.[45] on these two phonon modes). The remaining parameters adopted in our calculation are shown in Table III in Appendix C, unless otherwise stated. Notice, that the calculated trion binding energies for all five TMDCs studied fall in the range of 20-35 meV (see table I), in reasonable agreement with recent measurements [18, 41, 46-48]. These values are nearly an order of magnitude larger than that found in conventional quasi-2D systems, such as GaAs semiconductor quantum wells (QWs) [11, 12, 14, 18, 46, 47]. We note also that the similarity of trion binding energies in MoSe$_2$ and WSe$_2$ is perfectly reproduced in our calculation [18, 41, 48].

The results show also that in selenides, the trion binding energy matches one (for MoSe$_2$) or two (for WSe$_2$) phonon modes. This leads to strong coupling of the exciton and trion by exchange of an optical phonon, as discussed below [17, 20, 24, 26, 40, 49-51]. In sulfides, the calculated trion binding energies are sizeably smaller than both phonon energies [40, 49-51]. However, in these materials the measured trion-exciton energy separation may strongly depend on doping, as shown in ref [41, 44, 50] (for WS$_2$) and [41, 42, 50] (for MoS$_2$). Indeed, in sulfides, due to the presence of the 2D electron concentration (due to the intrinsic doping) the trion radiative recombination is possibly accompanied by the excitation of electrons of the Fermi gaz, leading to a higher energy separation between the peaks associated with the exciton and trion [41, 42, 44, 50]. For exemple, in a WS$_2$ un gated sample, with intrinsic doping $n_e = 1.710^{12} cm^{-2}$, the energy separation of the X and $X_-$ peaks in PL spectra ($\Delta E$) amounts to 45 meV, thus 11 meV higher than our result in table I, an energy difference in agreement with the Fermi energy of the electron gas in the sample [41, 44, 50]. Interestingly, this 45 meV energy separation is again nearly resonant with the energy of $E'$ optical phonon, a feature that may explain the efficient up- conversion PL process reported in ref.[24]. Recently, the family of semiconducting TMDCs monolayers expanded with a fifth binary material: MoTe$_2$ [19, 52, 53]. Using our models (see Appendix A), for MoTe$_2$ ML deposited in SiO$_2$ substrate and exposed to the air, the energy of the exciton and trion are equal to 1184.8 meV and 1157 meV respectively, yielding a trion binding energy of $\sim 27.8$ meV which roughly matches the $E'$ optical phonon energy in this material. Note finally that experiments show that like in selenides, the $X$-$X_-$ measured energy separation for MoTe$_2$ gives the trion binding energy, which is thus not affected by carrier doping effects as reported for sulfides. The study of such carrier concentration effects is beyond the scope of this work. Here below, we focus on the up and down (section III) conversion processes in selenides and MoTe$_2$ materials [19, 52, 53].

TABLE I. The energy of exciton $E_X$ and trion $E_{X_-}$ as well as its binding energy $E_{BX_-}$ for monolayer TMDCs exposed to the air and deposed in SiO2 substrates, $\hbar\omega_{HP}$ and $\hbar\omega_{LO}$ phonon are, respectively the energies of the homopolar and LO phonons.

|  | WSe$_2$ | MoSe$_2$ | WS$_2$ | MoS$_2$ | MoTe2 |
| --- | --- | --- | --- | --- | --- |
| $E_X$ (meV) | 1750 | 1640 | 2090 | 1870 | 1184.8 |
| $E_{X_-}$ (meV) | 1719 | 1610 | 2056 | 1841 | 1157 |
| $E_{BX_-}$ (meV) | 31 | 30 | 34 | 29 | 27.8 |
| $\hbar\omega_{LO}$ (meV) [40,49,51,54,55] | 31 | 36.95 | 44.14 | 46.33 | 27.72 |
| $\hbar\omega_{HP}$ (meV) [40,49,51,54,55] | 31 | 29.76 | 52 | 49.21 | 20.20 |

**2. Trion-optical phonon interactions**

The interaction between the trion and the optical phonons ($A_1'$,$E'$) is given by :

$$H^\lambda_{X_--ph} = \sum_Q \beta_j V^j_\lambda(Q) \ e^{-iQ\rho_j} \ \hat{a}^+_{\lambda,-Q} + c.c \quad (3)$$

where, $\hat{a}^+_{\lambda,-Q}(\hat{a}_{\lambda,Q})$ are the phonon creation (annihilation) operator for mode $\lambda$ = LO or HP with wave vector **Q**. The sum over $j$ accounts for the different particles (j= e$_1$, e$_2$, h). $\beta_j = 1$ ($\beta_j = -1$) when the *jth* particle is an electron (hole). $\rho_j$ is the in-plane position vectors for the *jth* particle and c.c. stands for the hermitian conjugate. The $A_1'$ mode corresponds to thickness fluctuations while the $E'$ mode corresponds to in-phase motion of the chalcogen atoms in the plane of the layer [17, 40, 49, 51, 54-60]. The insets of Fig.1 (a) show schematically their corresponding atomic displacements. The coupling of electrons to homopolar phonons, which corresponds to the irreducible representation $A_1^{'}$ of the symmetry group $D_{3h}$ is governed by the short-range potential induced by the volume change of the unit-cell volume. In polar crystals such as TMDCs, longitudinal optical (LO) phonons which corresponds to the irreducible representation $E^{'}$ of the symmetry group $D_{3h}$ of the crystal, induce a macroscopic electric field which couple to electrons and hole via long-range Frohlich interaction (FI) [17, 40, 49, 51, 54-60]. Appendix C.1-2 gives the expressions of $V_{HP}^j(Q)$, $V_{LO}^j(Q)$ and lists all the parameters we use.

3. Initial and final states in the up-conversion process

To derive the electron-phonon matrix element $\langle \psi_F | H_{X_-ph}^\lambda | \psi_I \rangle$ in Eq.(1), we need to introduce the initial and final states (wave functions, energies) for the up-conversion process. As shown in Appendix A, the trion wave function is conveniently described in terms of the centre of mass $M_{X_-}\boldsymbol{R}_{X_-} = m_e(\boldsymbol{\rho}_{e1} + \boldsymbol{\rho}_{e2}) + m_h\boldsymbol{\rho}_h$ and the two relative $\boldsymbol{\rho}_n = \boldsymbol{\rho}_{en} - \boldsymbol{\rho}_h$ coordinates. One has for the initial trion states:

$$\Psi_{X_-}^i(\boldsymbol{R}_{X_-}, \boldsymbol{\rho}_1, \boldsymbol{\rho}_2) = \frac{1}{\sqrt{A}} e^{i\boldsymbol{K}_{X_-}^i \boldsymbol{R}_{X_-}} \boldsymbol{\zeta}_{X_-}(\boldsymbol{\rho}_1, \boldsymbol{\rho}_2) \quad (4)$$

The corresponding energy can be written as :

$$E_{X_-}^i = E_{X_-} + \frac{\hbar^2 {K_{X_-}^i}^2}{2M_{X_-}} \quad (5)$$

The eigenvalues $E_{X_-}$ and the eigenfunctions $\boldsymbol{\zeta}_{X_-}(\boldsymbol{\rho}_1, \boldsymbol{\rho}_2)$ are obtained numerically by a direct diagonalization of the trion relative Hamiltonian matrix (see Appendix A ). $M_{X_-} = 2m_e + m_h$ is the trion mass, $\boldsymbol{K}_{X_-}^i$ is the wavevector related to the free centre-of- mass propagation and A is the two-dimensional quantization area in the monolayer plane. The total trion-phonon states are then given by :

$$|\psi_I\rangle_{K_{X_-}^i, Q} = |\Psi_{X_-}^i\rangle_{K_{X_-}^i} \otimes |N_{\lambda,Q}\rangle \quad (6)$$

with energy:

$$E_I(\boldsymbol{K}_{X_-}^i, \boldsymbol{Q}) = E_{X_-}^i + \hbar\omega_{\lambda,Q} N_{\lambda,Q} \quad (7)$$

where, $|N_{\lambda,Q}\rangle$ is the number state of the phonon mode $\lambda$ with wave vector $\boldsymbol{Q}$. The thermalized average occupation number of the phonons is given by the Bose distribution $N_{\lambda,Q} = \left[\exp\left(\frac{\hbar\omega_{\lambda,Q}}{K_B T}\right) - 1\right]^{-1}$, $\hbar\omega_{\lambda,Q}$ is the optical phonon energy that is absorbed during the up-conversion process. After absorbing a $\boldsymbol{Q}$-wave vector phonon from the bath, $X_-$ with wave vector $\boldsymbol{K}_{X_-}^i$ makes a transition towards a dissociated state corresponding to an exciton $X$ with center of mass wave vector $\boldsymbol{K}_X$ and one unbound electron with wave vector $\boldsymbol{q}_e$. As shown in Appendix A , the two wavevectors $\boldsymbol{K}_{X_-}^f = \boldsymbol{K}_X + \boldsymbol{q}_e$ (describing the free centre-of-mass propagation) and $\boldsymbol{K} = \frac{m_e\boldsymbol{K}_X - M_X\boldsymbol{q}_e}{M_{X_-}}$ (related to the relative motion of the dissociated exciton-electron pair) permit an equivalent description of the final trion state. Note that the two sets of wavevectors are related by a unit-jacobian transformation. We shall use the latter below. The total final states thus read:

$$|\psi_F\rangle_{K_{X_-}^f,K,Q} = |\Psi_{X-e}\rangle_{K_{X_-}^f,K} \otimes |N_{\lambda,Q} - 1\rangle \quad (8)$$

with energy:

$$E_F(K_{X_-}^f, K, Q) = E_X + \frac{\hbar^2 {K_{X_-}^f}^2}{2M_{X_-}} + \frac{\hbar^2 K^2}{2\mu} + \hbar\omega_{\lambda,Q}(N_{\lambda,Q} - 1) \quad (9)$$

where $E_X$ is the exciton eigenenergie solution of the exciton relative hamiltonian $H_X$ (see Appendix A) and $\mu = \frac{m_e m_h}{M_X}$ is the electron-exciton reduced mass. The wave function of the unbound electron-exciton is given as follow :

$$\Psi_{X-e}(R_{X_-}, \rho_1, \rho_2) = \frac{1}{\sqrt{A}} e^{iK_{X_-}^f \cdot R_{X_-}} \zeta_{X-e}(\rho_1, \rho_2) \quad (10)$$

where,

$$\zeta_{X-e}(\rho_1, \rho_2) = \frac{1}{\sqrt{2A}} \left[ e^{-iK(\rho_1 - \beta_X \rho_2)} \phi_{\widetilde{1s}}(\rho_2) + e^{-iK(\rho_2 - \beta_X \rho_1)} \phi_{\widetilde{1s}}(\rho_1) \right] \quad (11)$$

here, $\beta_X = \frac{m_e}{M_X}$. Momentum conservation during the transition puts no constraints on the relative momentum $K$, so there is a continuum of final states, justifying thereby the use of the Fermi golden rule to evaluate in the following the up-conversion rate.

## 4. Trion to exciton up-conversion rate

Finally, after introducing the final and initial states, the phonon mode involved in the scattering process as well as the trion-phonon interaction operator, we are ready now to investigate trion to exciton up conversion rate given in Eq.(1). Now inserting Eqs.(3), (6) and (8) into Eq.(2), the matrix element describing the up-conversion process from trion state with the center of mass wave vector $K_{X_-}^i$ to one unbound electron-exciton state with the center of mass wave vector $K_{X_-}^f$ and relative wave vector $K$ assisted by absorption of optical phonon ($\lambda$= LO or HP ) is given by :

$$F_\lambda^{UC}(Q, K, K_{X_-}^f, K_{X_-}^i) = \langle \Psi_F | H_{X_- - ph}^\lambda | \Psi_I \rangle = \sqrt{N_{\lambda,Q}} \times \delta_{K_{X_-}^f - (K_{X_-}^i + Q)} \mathbb{M}_{fi}(Q, K) \quad (12)$$

Where

$$\mathbb{M}_{fi}(Q, K) = \left[ V_\lambda^{e_1}(Q) \mathbb{I}_{e_1}(Q, K) + V_\lambda^{e_2}(Q) \mathbb{I}_{e_2}(Q, K) - V_\lambda^h(Q) \mathbb{I}_h(Q, K) \right] \quad (13)$$

where, the matrix element that describes the interaction of electrons and hole with polar optical phonons is given, respectively, by :

$$\mathbb{I}_{e_1}(Q, K) = \langle \zeta_{X-e}(\rho_1, \rho_2) | e^{iQ(\beta_{X_-} \rho_1 - \gamma_{X_-} \rho_2)} | \zeta_{X_-}(\rho_1, \rho_2) \rangle \quad (14)$$

$$\mathbb{I}_h(Q, K) = \langle \zeta_{X-e}(\rho_1, \rho_2) | e^{-i\gamma_{X_-} Q(\rho_1 + \rho_2)} | \zeta_{X_-}(\rho_1, \rho_2) \rangle \quad (15)$$

$\mathbb{I}_{e_2}(\boldsymbol{Q},\boldsymbol{K})$ is obtained just by replacing $\rho_1 \leftrightarrow \rho_2$ in the exponential term of the Eq.(14). In the above equations $\beta_{X_-} = \frac{M_X}{M_{X_-}}$, and $\gamma_{X_-} = \frac{m_e}{M_{X_-}}$, for more detail see Appendix C.3. Translational invariance along the ML implies the conservation of the in-plane momentum $|K_{X_-}^f| = \sqrt{K_{X_-}^{i\,2} + Q^2 + 2K_{X_-}^i Q \cos\theta_{K_{X_-}^i,Q}}$ as explicitly shown in Eq.(12) by the presence of the Kronecker $\delta$. Now, inserting Eq.(12) into Eq.(1), taking into account properties of Kronecker $\delta$ and changing the sum over $\boldsymbol{Q}$ and $\boldsymbol{K}$ into integrals, the up conversion rate for a bound trion with fixed initial centre-of-mass wavevector $K_{X_-}^i$ can be rewriten as follow:

$$W_{\lambda, X_- \to X}^{UC}(\boldsymbol{K}_{X_-}^i) = \frac{1}{\tau_{\lambda, X_- \to X}^{UC}} = \frac{2\pi}{\hbar} \frac{B_\lambda(T)}{(2\pi)^4} \int_0^\infty Q\, dQ \int_0^{2\pi} d\theta_Q \int_0^{2\pi} d\theta_K \left|\mathbb{M}_{fi}(Q, K_0)\right|^2 \quad (16)$$

Where $B_\lambda(T) = \frac{N_\lambda A^2 \mu}{\hbar^2}$. In Eq.(16), $K_0 = \sqrt{\frac{2\mu}{\hbar^2}(E_{X_-} - E_X + \hbar\omega_\lambda) - \frac{\mu K_{X_-}^{i\,2}}{M_{X_-}} + \frac{\mu K_{X_-}^{f\,2}}{M_{X_-}}}$ is the root of the argument of the delta function Eq.(1). The integration is taken over the range in which terms under the square root are positive. Due to the dispersionless nature of long-wavelength optical phonons, we have neglected its weak dependence on Q i.e. $\omega_{\lambda,\boldsymbol{Q}} \sim \omega_\lambda$.

**B. Numerical results of the $X_-$ - $X$ up-conversion rate**

In Fig.1(b), we display the trion to exciton up conversion rate for WSe$_2$ monolayer, assisted by both homopolar $A_1'$ and LO $E'$ optical phonons, as a function of $K_{X_-}^i a_b$ for fixed temperature T=100 K, where $a_b$ is the Bohr radius of the exciton. The results show that the scattering rate is significant for zero initial trion wave vector $K_{X_-}^i$, decreases rapidly with increasing $K_{X_-}^i$ for both phonon couplings. The largest scattering probability for $\boldsymbol{K}_{X_-}^i = \boldsymbol{0}$ can be justified by the $\frac{1}{(cte + a_b^2|\alpha Q \pm \beta K|^2)^{3/2}}$ dependence of the matrix element $\mathbb{I}_i(\boldsymbol{Q},\boldsymbol{K})$ discussed in Appendix C.3. This behavior implies an efficient population transfer between $X_-$ and $X$ for zero intitial trion wave vector. We can notice also that the up-conversion rate strongly depend on the phonon mode.

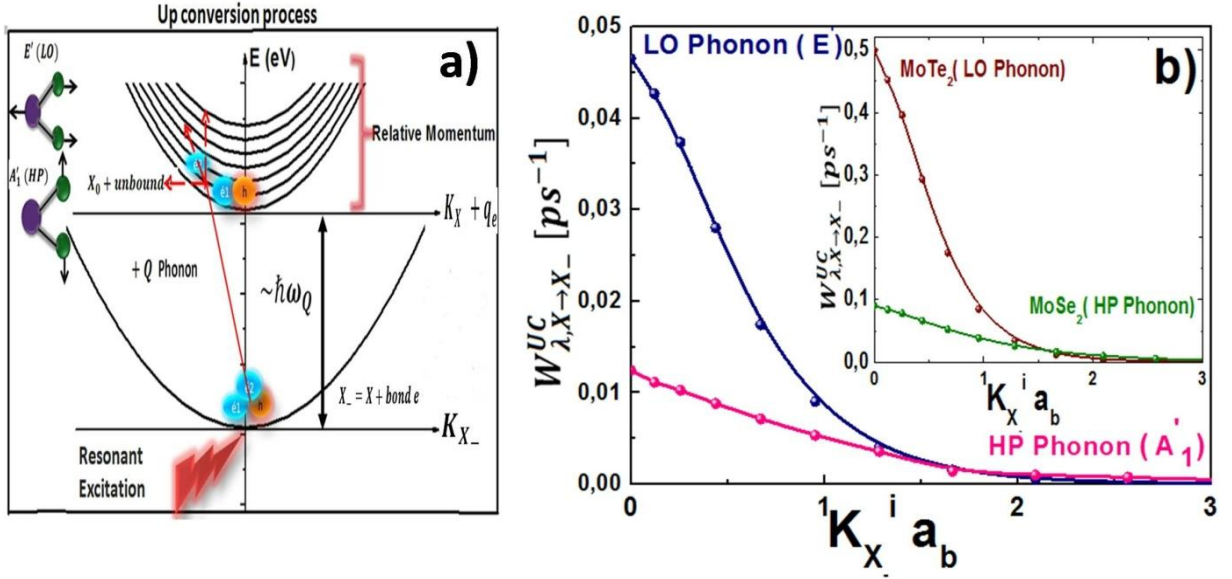

FIG. 1. a) A scheme of the negatively charged and neutral exciton energy level diagram. The lowest curve represents the trion dispersion with its centre-of-mass momentum $K_{X_-}$. The highest curves represent the dispersions of the unbound exciton-electron pair with their centre-of-mass momentum $K_X + q_e$, different curves correspond to different relative momenta so they form an energy continuum. By absorbing a phonon with momentum $Q$, the trion can be converted into an unbound exciton-electron pair. The inset shows a picture of the atomic displacements of phonons $A_1'$ and $E'$. b) Trion to exciton up-conversion rate due to homopolar (pink line) and LO phonon (blue line) for WSe$_2$ deposited on SiO$_2$ substrate with high dielectric frequency $\epsilon_b = 2.1$ as a function of $K_{X_-}^i a_b$ for T = 100K, the insert shows the trion to exciton up-conversion rate in various monolayer system MoSe$_2$ ( green line), and MoTe$_2$ ( brown line).

In fact, comparing the two curves, we see that the optical LO phonon dominates the anti-Stokes emission over the HP phonon. Calculating the thermal average up-conversion rate,

$$\langle W_{\lambda,X_-\to X}^{UC} \rangle = \langle \frac{1}{\tau_{\lambda,X_-\to X}^{UC}} \rangle = \frac{\sum_{K_{X_-}^i} f(K_{X_-}^i) W_{\lambda,X_-\to X}^{UC}((K_{X_-}^i)}{\sum_{K_{X_-}^i} f(K_{X_-}^i)} \quad (17)$$

at T=100 K, we find that the thermally averaged up-conversion time via $E'$ phonon (∼33ps ) is about 4 times faster than the one calculated in the case of the coupling with $A_1'$ phonon , ∼ 118ps, see table II. This conversion time is in a agreement with the experimental results of Jones *et al.*[17], in which they measured, at the same temperature, an $A_1'$ phonon assisted up- convertion time in the order of 100 ps. The slightly difference can be due to the parameters choosen in our calculation. Notably, the trion distribution function $f(K_{X_-}^i)$ can be approximated, in the low density limit, by the Boltzmann distribution $f(K_{X_-}^i) = exp\left(-\frac{\hbar^2 {K_{X_-}^i}^2}{2M_{X_-} K_B T}\right)$. We discuss in section (B.1) the influence of the substrate dielectric screening and in section (B.2) the role of temperature on the up-conversion rates in WSe$_2$. Finally, in section(B.3) we extend our study to other monolayer TMDCs.

# 1. Dielectric environment dependence of the thermally averaged up-conversion rate

The up-conversion rate of the HP phonon is independent of the dielectric environment, as its coupling mechanism does not involve an electric field [51, 54, 55, 60] while for the LO phonon, the strength of Frohlich interaction depends on the polarization properties of the outside materials [51, 54, 55, 60]. In fact, it has been shown that screening plays a fundamental role in the phonon-momentum dependency of the polar-optical coupling. This effect can be associated with the formation of surface charges due to the change in the dielectric properties at the interfaces between the 2D material and its environment [54, 55]. As shown in Fig.2, it is clear that dielectric screening is of paramount importance to evaluate the strength of the Frohlich interaction and hence the decay rate of population from the initial $X_-$ state to the $X$-electron continuum. In fact, an important decrease (increase) of the thermally averaged up-conversion rate $\langle W_{LO,X_- \to X}^{UC} \rangle$ (the thermally averaged up-conversion time) is observed when increasing either the average dielectric environment constant $\varepsilon_{eff}$ and the screening length $r_{eff}$ (see Eq.C.2 in Appendix C for details ).

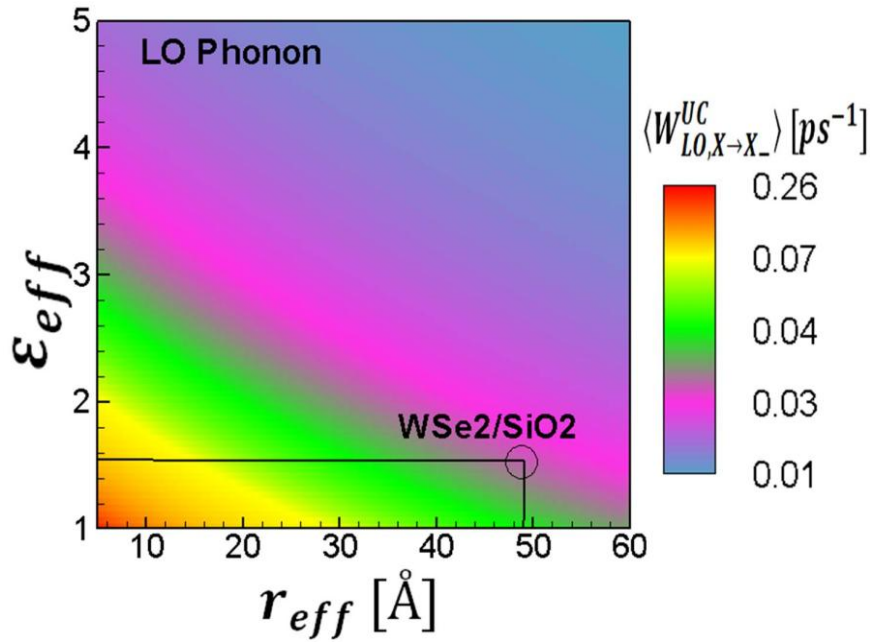

FIG. 2. Effect of the dielectric environment: LO phonon assisted trion to exciton thermal average up-conversion rate as a function of the the average dielectric constant of the surrounding material $\epsilon_{eff}$ and of the screening length $r_{eff}$ for T=100 K.

It is noteworthy that, for fixed value of $r_{eff}$ ( here for WSe$_2$ $r_{eff} = 48.7$Å) and without environmental screening (suspended monolayer) the $X_-$ - X thermally averaged up-conversion time is about 20 ps, while it rises to 67 ps when deposited on Micca substrate with dielectric constant $\varepsilon_b = 5$, and reaches 111 ps when the monolayer is encapsulated between the hBN akes, which is comparable to the up conversion rate calculated in the case of the HP phonon. Actually, the introduction of a dielectric environment through a substrate or full encapsulation leads to the reduction of the electric potential induced by the LO phonon in the long wavelength, by the dielectric constant of the environment, which influences strongly the efficiency of the exciton to trion up conversion rate (see Fig.2) [54, 55, 61].Notably, the rates due to LO and HP phonon coupling , become comparable for sufficiently large $r_{eff}$ values . It is however important to bear on mind that the model applies only for a system consisting of a thin material sheet with high dielectric constant as compared to those of the surrounding medium.

## 2. Temperature dependence of the thermally averaged up-conversion rate

Fig.3 shows the trion-to exciton thermally averaged up-conversion rate (in the inset of fig 3 the conversion rate as a function of $K_{X_-}^i a_b$) due to both $E'$ and $A_1'$ phonons, for WSe$_2$ monolayer over a temperature range of 5-200 K. We observe a clear increase of the thermally averaged conversion rate when raising the temperature, which is related to the increasing probability for phonon absorption due to the increasing occupation probability of both $A_1'$ and $E'$ optical phonon modes with increased thermal energy. Notably, TMDCs materials are characterized by polar optical-phonon energy $\hbar\omega_\lambda$ much greater than the thermal temperature $K_B T$ even at room temperature (see table.I), except in the case of the homopolar $A_1'$ phonon in MoTe$_2$ which is about 20 meV. In fact, in the low temperature range ($T \leq 50\ K$) the population transfer from $X_-$ to $X$ is completely blocked. The up- conversion time decreases then rapidly with increasing temperature : ~ 0.99 ns (~11 ps) for LO phonon and ~3.9 ns (~ 37 ps) for HP phonon at $T = 50$ K (150 K). At room temperature the up-conversion becomes very efficient, with sub-ps characteristic times.

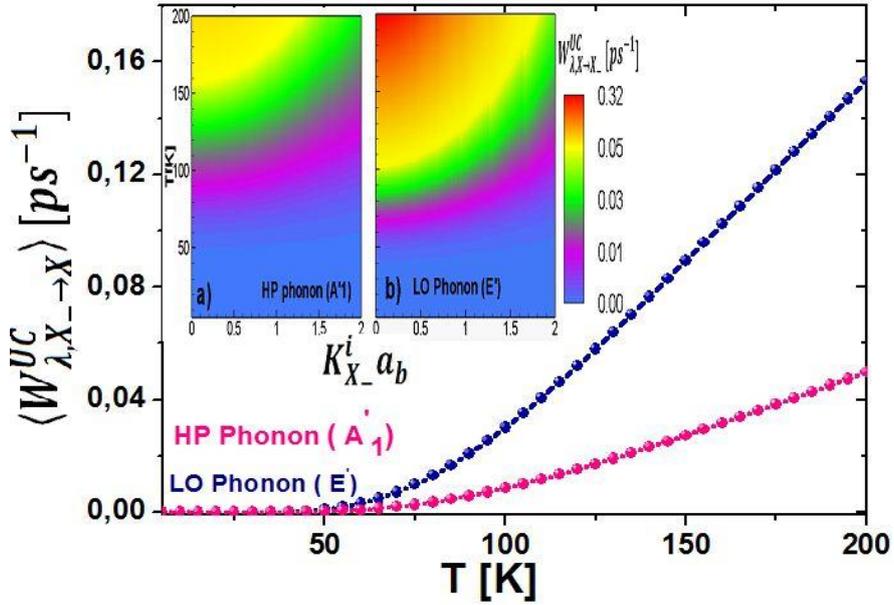

FIG. 3. Temperature dependence of the trion to exciton thermal average upconversion rate $\langle W_{\lambda,X_-\to X}^{UC}\rangle$ due to $A_1'$ homopolar phonon (pink curve) $E'$ LO phonon (blue curve), for WSe$_2$ deposited on SiO$_2$ substrate with high dielectric frequency $\epsilon_b = 2.1$ and exposed to the air $\epsilon_t = 1$. In the inset: the up-conversion rate $W_{\lambda,X_-\to X}^{UC}(K_{X_-}^i)$ due to a) $A_1'$ phonon b) $E'$ phonon.

The efficiency of the up conversion process at high temperature can explain the observed temperature variations of the trion and exciton integrated PL intensities in the selenides [61]. In selenides, the photoluminescence intensity of a trion is high at low temperatures. As the temperature is increased the relative luminescence of charged vs. neutral exciton decreases, with the neutral exciton eventually becoming the dominant feature [41, 43,

44, 50, 61, 62]. In fact, it has been shown that the trion integrated PL intensity is totally suppressed at about 150 K [41, 43, 44, 50, 62] (in some group at about 125 K [61]). Moreover, Robert *et al.*[61] observe that the trion (exciton) PL decay time decreases (increases) strongly when T increases. This is consistent with our results which predict that the up conversion becomes an important additional channel for trion population decay above 150 K.

**3. Trion to exciton up conversion rate for other monolayer TMDCs**

The theoretical results obtained in this study for WSe2 are applicable to other low dimensional transition metal dichalcogenides. The inset of Fig.1(b) shows that the up conversion rate $W_{\lambda,X_-\to X}^{UC}(K_{X_-}^i)$ is strongly dependent upon the materials and for a better comparison with available experimental measurements, table II summarizes the thermally averaged up-conversion time $\frac{1}{\langle W_{\lambda,X_-\to X}^{UC}\rangle}$ for different materials. Since we demonstrate in section (A.1) that the trion binding energy of MoSe$_2$ is nearly resonant with just $A_1'$ phonon, we plot in the inset of Fig.1(b) the up conversion rate of MoSe$_2$ assisted by A01 phonon (green line). We see that MoSe$_2$ has significantly higher up conversion rate than WSe2. Actually, the high HP deformation potentials, high exciton effective masses, low mass densities in MoSe2 as compared to that of WSe$_2$, and the comparable value of the HP phonon energy in both materials, result in stronger electron-phonon coupling and hence shorter thermal average up-conversion time $\frac{1}{\langle W_{HP,X_-\to X}^{UC}\rangle} = 15$ps , which is about 8 times faster than the one associated to WSe$_2$ (see also table II). Notably, this value is in a agreement with the experimental results of Kai Hao *el al.*[26] in which they deduce an up-conversion time for MoSe2 about 10 ps.

TABLE II. Thermal average up-conversion times for different TMDCs materials deposited on SiO2 substrate and exposed to the air at T=100 K.

|  | HP phonon | | LO phonon | |
|---|---|---|---|---|
| TMDCs materials | MoSe$_2$ | WSe$_2$ | WSe$_2$ | MoTe$_2$ |
| $\langle\frac{1}{\tau_{\lambda,X_-\to X}^{UC}}\rangle$ | 15 ps | 118 ps | 33 ps | 3.25 ps |

In the same figure, we plot also the up conversion rate for MoTe$_2$ ( Brown insert) assisted by $E'$ phonon mode. We clearly notice that MoTe2 has the shortest thermal average conversion time $\frac{1}{\langle W_{LO,X_-\to X}^{UC}\rangle}$ = 3.25ps , due to the higher Frohlich interaction strength and lower LO phonon energy. Finally, the inset of Fig.1(b) also indicates that the up-conversion process in all the TMDCs monolayers studied in our work are more probable for zero center-of-mass wavevectors .

# III. PHONON ASSISTED EXCITON TO TRION DOWN-CONVERSION PROCESS

Alternatively, exciton to trion down-conversion process is a spontaneous emission, which concerns the appearance of a low-energy (Stokes) PL after the formation of a trion from a dissociated exciton-electron pair in a process trigered by the emission of a phonon. In our work, we assume an initial gas (i.e. before exciton photo-generation) of $N_e$ free electrons with energy $E(eB_i) = \sum_{q_e} \frac{\hbar^2 q_e^2}{2m_e}$, where $|eB_i\rangle$ is the initial electron bath. Hence, when the $X$ state is resonantly excited, the generated exciton can combine with a free electron to become $X_-$, accompanied by the emission of a phonon. We assume also (like for the up-conversion case) a model of non-interacting gas- exciton particles. In this case, we can consider any two (called "1" and "2" here below) of the $N_e + 1$ electrons to write the wavefunction of the dissociated trion as :

$$|\psi_I\rangle_{K_X,q_e,Q} = |\Psi_{X-e}\rangle_{K_X,q_e} \otimes |N_{\lambda,Q}\rangle \quad (18)$$

with energy :

$$E_I(K_X, q_e) = E_X + \frac{\hbar^2 K_X^2}{2M_X} + \frac{\hbar^2 q_e^2}{2m_e} + E_i^{ph} \quad (19)$$

here, the electron-exciton wave function $\Psi_{X-e}(R_{X_1}, R_{X_2}, \rho_{e1}, \rho_{e2})$ is given by Eq.(A.4) in Appendix A. For the final state one has under the same assumptions:

$$|\psi_F\rangle_{K_{X_-}^f,Q} = |\Psi_{X_-}^f\rangle_{K_{X_-}^f} \otimes |N_{\lambda,Q} + 1\rangle \quad (20)$$

with energy :

$$E_F(, K_{X_-}^f, Q) = E_{X_-} + \frac{\hbar^2 K_{X_-}^{f\,2}}{2M_X} + E_i^{ph} + \hbar\omega_\lambda \quad (21)$$

$\Psi_{X_-}^f(R_{X_-}, \rho_1, \rho_2)$ is the trion wave function. Finally, the interaction of the phonon reservoir with the three particles systems is given by the same expression as for the up-conversion process. We obtain thus the down-conversion rate for a given initial $K_X$ value :

$$W_{\lambda,X \to X_-}^{DC}(K_X) = \frac{2\pi}{\hbar} \sum_{Q q_e, K_X} |F_\lambda^{DC}(Q, q_e, K_X, K_{X_-}^f)|^2 f_e(q_e)\, \delta\left(E_F(K_{X_-}^f, Q) - E_I(K_X, q_e)\right) \quad (22)$$

Where

$$F_\lambda^{DC}(Q, q_e, K_X, K_{X_-}^f) = \sqrt{N_\lambda + 1}\, \delta_{Q+K_X+q_e, K_{X_-}^f}\, \mathbb{M}_{fi}(Q, K_X, q_e) \quad (23)$$

The occupation numbers $f_e(q_e)$ of electron states in Eq.(22) can be modeled using the Fermi-Dirac distribution: $f_e(q_e) = \left(e^{(\frac{\hbar^2 q_e^2}{2m_e} - \mu_e)/K_B T} + 1\right)^{-1}$, where the chemical potential $\mu_e$ is given by $\mu_e = K_B T \ln\left(e^{\frac{N_e}{g_{2D} K_B T}} - 1\right)$, with $N_e$ is the electron two-dimensional density and $g_{2D}$ is the 2D density of states. Here, we assume that the electrons and phonons have equilibrated to a common temperature $T_e = T_l = T$. We note finally that taking into consideration from the beginning antisymetrized contributions involving the ensemble of $N_e + 1$ electrons (e.g. by using Slater determinants) leads to the same expression for the scattering rate, as expected for an ensemble of non-interacting electrons. Fig.4(a) shows the calculated down-conversion rate as a function of $K_X a_b$ with emission of an LO phonon (blue line) and HP phonon (pink line), for a fixed temperature T = 100K and electron density $N_e = 10^{12} cm^{-2}$. Similar to the up-conversion process, down-conversion process occurs most rapidly at zero exciton center-of-mass wavevectors, then it fastly vanish at $K_X > 0.076 Å^{-1}$. Comparing the up and down-conversion times at zero center of mass wave vector, we notice that the time required for $X \rightarrow X_-$ down conversion process is about 15 times shorter ($\tau_{LO,X \rightarrow X_-}^{DC} = 1.38$ps, $\tau_{HP,X \rightarrow X_-}^{DC} = 5.37$ps ) than $X_- \rightarrow X$ up-conversion time (respectively around 22 ps and 83 ps, see Fig.1(b)). This is half the value expected from the simple phonon population factor $\frac{N_Q + 1}{N_Q} = 30$ for this material at T=100 K. Indeed, the two processes probe different final density of states and, most importantly, the down-convertion depends also upon the free gas density, as is clear from its dependence upon the free gas chemical potential (see discussion below). As we mentioned before, the excitons are photogenerated resonantly, hence the wavevector of exciton states formed due to optical excitation lies close to zero due to selection rules. Actually, the lifetime of the photocreated exciton at $K_X = 0$ is very short (in the picosecond time scale) and since the excitons density is not very large, as we assumed in our work, then the exciton-exciton and exciton acoustic phonon interaction may not be very effective for the the thermalization process. In the following since the down-converssion process efficient at zero center of mass wave vector, so we choose $K_X = 0$ as the limitation value for all numerical calculations presented in this section.

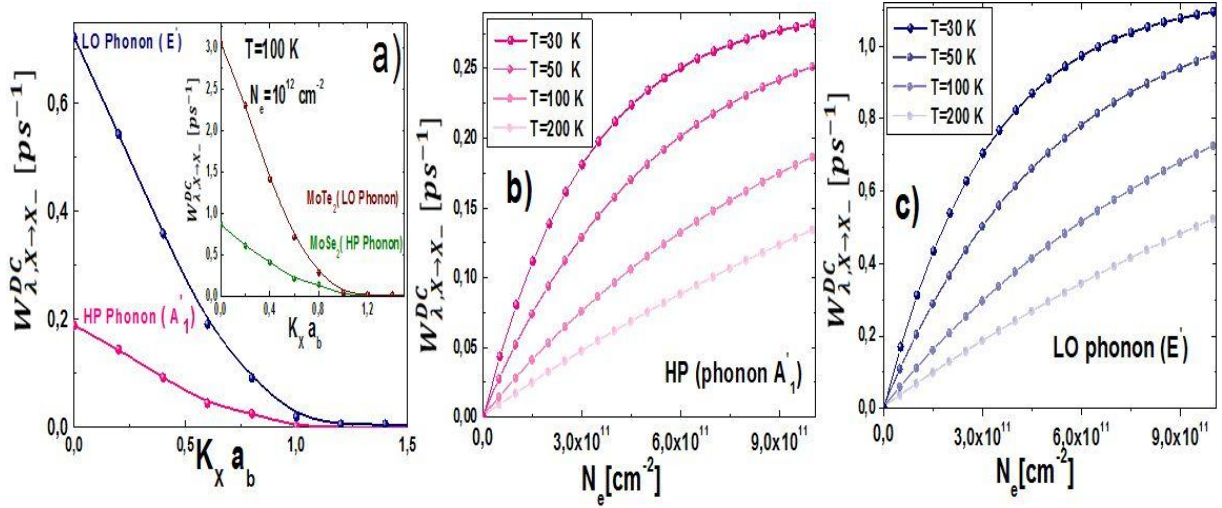

FIG. 4. a) Exciton to trion down-conversion rate due to homopolar (pink line) and LO phonon (blue line) for WSe2 deposited on SiO2 substrate with high dielectric frequency $\epsilon_b = 2.1$ as a function of $K_X a_b$ for T=100 K and electron density $N_e = 10^{12}\, cm^{-2}$. The insert show down-conversion rate as a function of $K_X a_b$ for MoSe2 (green line), and MoTe2 ( brown line). The down-conversion rate, for $K_X = 0$ as a function of the electron density $N_e$, at different temperatures (T = 30K; 50K; 100K; 200K for b) HP phonon and c) LO phonon in WSe2 monolayer

In Fig.4 (b,c), we show the electron density $N_e$ dependence of the down-conversion rate due to LO and HP phonon scattering at four differents temperatures (T = 30, 50, 100, 200 K). The results show that for a given T and for $N_e < 10^{10}\, cm^{-2}$, the exciton to trion down-conversion process is very unlikely, as the conversion rate becomes much longer than the exciton radiative lifetime, i.e., the neutral exciton would have already decayed before capture of one electron. For exemple, for $N_e = 10^9\, cm^{-2}$, at $T = 10K$, we find $\tau^{DC}_{HP,X\rightarrow X_-} = 358$ps, and $\tau^{DC}_{LO,X\rightarrow X_-} = 92$ps (see Appendix D) which is much more slower than the exciton radiative rate $\tau^r_X = 4$ps (the exciton radiative lifetime is calculated using Eq.(E.1-2) in Apeendix E). When the electron density increases, the down conversion-rate increases dramatically and can reach a few picoseconds time scale. Using the numerical values of the down-conversion rate, we performed numerical fits using the following relation which involve the carrier concentrations:

$$W^{DC}_{\lambda,X\rightarrow X_-} = W^{DC}_{0,\lambda}\left(1 - e^{\frac{N_e}{N_0}}\right) \quad (24)$$

where $W^{DC}_{0,X\rightarrow X_-}$ and $N_0$ are the fitting parameters. For the curves in Fig.4(b,c), we get $W^{DC}_{0,HP} = 0.31$ps for HP phonon and $W^{DC}_{0,LO} = 1.2$ps for the 4 temperatures, and $N_0 = 3.14 10^{11}\, cm^{-2}$ for T=30 K, $N_0 = 5.23 10^{11}\, cm^{-2}$ for T=50 K, $N_0 = 1 10^{12}\, cm^{-2}$ for T=100 K and $N_0 = 2 10^{12}\, cm^{-2}$ for T=200 K for both phonon mode (HP and LO).It is finally worth noticing that the down conversion rate decreases with increasing temperature, for a given gas density. This results mostly from the fact that electrons acquire greater wavevectors, leading to a sizeable decrease of the phonon matrix elements.

The increasing of down conversion rate with Ne leading to relative enhancement of trion PL especially at low temperature (T < 50K) since in this temperature regime the up-conversion process (exciton formation) is blocked while the down conversion process (trion formation) become more efficient ( faster down conversion time with decreasing T) (see also Appendix D). It is noteworthy that our numerical results for the temperature and electron density dependence of the down and up conversion process due to LO and HP scattering is consistent with experimental finding of Robert et al .[61].

In the insert of fig 4.(a), we calculate the down conversion rate, for different TMDCs materials. Similar to up conversion process, there are subtle causes to the differences obtained in the down conversion rate due to varying material properties of the different monolayer systems. For $N_e = 10^{12} cm^{-2}$ and T=100 K, MoTe2 possesses the fastest down conversion time, in the order of sub-picosecond time scale ($\tau_{LO,X \to X_-}^{DC} = 0.32$ps), yielding an efficient exciton to trion population transfer, since it is about one order of magnitude shorter than the exciton radiative time . For the HP phonon, the results show that MoSe$_2$ (green line) possesses a down conversion time about 5 times faster than WSe2 (pink line), in the order of 1 ps, see Appendix D for more detail of the variation of down-conversion rate of MoSe$_2$ and MoTe$_2$ as a function of T and $N_e$.

## IV. CONCLUSION

The coupling between phonons and charged particles in monolayer transition-metal dichalcogenides (ML-TMDs) exhibits unique behavior due to their atomically thin nature. In our study, we show that selenides (MoSe$_2$, WSe$_2$) and MoTe$_2$ materials, possess charged exciton binding energies nearly resonant with corresponding optical phonon energies ($A_1'$ and/or $E'$), and that this energy proximity leads to an efficient population transfer between exciton and trion mediated by optical phonon. In this context, using a theoretical model based on the Fermi golden rule, we have calculated the phonon assisted trion (exciton) to exciton (trion) up (down) conversion rate. For WSe$_2$ momolayer deposited on the top of SiO$_2$ substrate and exposed to the air and for T=100 K, $N_e = 10^{12} cm^{-2}$ and zero initial center of mass wave vector, we obtained an $X \to X_-$ conversion time of few picoseconds time scale and a longer $X_- \to X$ up-conversion time, in the order of tens picoseconds, since this is an anti-Stokes scattering process. We have demonstrated that the efficiency of the up and down conversion process is strongly dependent upon the experimental condition such as: temperature, dielectric environment, electron density. In fact, we have found that the up conversion rate increases dramatically with increasing temperature due to the augmentation of the average phonon occupation $N_e$ with T, yielding an efficient population transfer from $X_-$ to X at high T. Conversely, one strikingly obtains that the down-conversion rate decreases with increasing temperature, a result due to the strong effect of temperature on the free electron gas distribution . For $N_e < 10^{10} cm^{-2}$, the down conversion process becomes very unlikely, since the process occurs on time scale much longer than the exciton lifetime. Notably, our results can explain the behavior of the trion and exciton integrated PL intensity, observed on experiment, as a function of T. We demonstrated also that while the conversion process due the HP phonon is independent of the dielectric environment, the transfer of population between $X_-$ and X is strongly influenced by any additional screening from the dielectric environment surrounding the monolayer. Finally, we have obtained that the conversion processes depend also to the choice of the materials, and we showed that MoTe$_2$ have the shortest up (2 ps) and down (0.3 ps) conversion time.

**Appendix A: TRION STATES**

In the presence of residual free charge carriers, charged excitons can be formed. The strong Coulomb interaction in monolayer TMDs leads to large trion binding energy [18, 47, 63]. We consider here negatively charged trions that consist of two electrons and one hole all of which reside in the same valley. In terms of center of mass coordinate $\boldsymbol{R}_{X_-}$ and relative coordinates of each electron with respect to the hole, $\boldsymbol{\rho}_1$ and $\boldsymbol{\rho}_2$, the Schrodinger equation in the effective mass approximation is given by:

$$\left(-\frac{\hbar^2 \nabla_{R_{X_-}}^2}{2M_{X_-}} + H_{X_-}^{rel}\right)\Psi_{X_-}^i(\boldsymbol{R}_{X_-},\boldsymbol{\rho}_1,\boldsymbol{\rho}_2) = E_{X_-}^i\ \Psi_{X_-}^i(\boldsymbol{R}_{X_-},\boldsymbol{\rho}_1,\boldsymbol{\rho}_2) \quad \text{(A.1)}$$

where the relative trion motion is described by :

$$H_{X_-}^{rel} = \sum_{i=1,2} H_{X_i} - \frac{\hbar^2}{m_h} \nabla_{\boldsymbol{\rho}_1}\cdot\nabla_{\boldsymbol{\rho}_2} - V_{2D}(|\boldsymbol{\rho}_1 - \boldsymbol{\rho}_2|) \quad \text{(A.2)}$$

here, $H_{X_i} = -\frac{\hbar^2 \nabla_{\rho_i}^2}{2\mu} + V_{2D}(\rho_i)$ is the relative Hamiltonian of the neutral exciton ( See ref [63] for more detail ), $\mu_X = \frac{m_e m_h}{M_X}$ is the reduced effective mass. $V_{2D}(\rho_i)$ is the nonlocally-screened electron hole interaction originating from the change in the dielectric environment [47, 63-65]. The total trion wave function solution of Eq.A1 can be written as the product of the trion center of mass contribution and the relative wave function: $\Psi_{X_-}^i(\boldsymbol{R}_{X_-},\boldsymbol{\rho}_1,\boldsymbol{\rho}_2) = \frac{1}{\sqrt{A}}\ e^{iK_{X_-}^i\cdot R_{X_-}}\zeta_{X_-}(\boldsymbol{\rho}_1,\boldsymbol{\rho}_2)$. The expressions of $\zeta_{X_-}(\boldsymbol{\rho}_1,\boldsymbol{\rho}_2)$ for the initial and final states are discussed below. Owing to the exchange of electrons, both singlet ($\zeta(\boldsymbol{\rho}_2,\boldsymbol{\rho}_1) = +\zeta(\boldsymbol{\rho}_1,\boldsymbol{\rho}_2)$) and triplet ($\zeta(\boldsymbol{\rho}_2,\boldsymbol{\rho}_1) = -\zeta(\boldsymbol{\rho}_1,\boldsymbol{\rho}_2)$) trion states may occur [66], for both the initial (bound trion) and final (dissociate trion) configurations. The existence of a trion fine structure (singlet and triplet states) in the PL spectra of TMDCs materials depends strongly on the type of the materials (Mo/W) [17, 41, 67-72]. It has been demonstrated theoretically and experimentally that the observation of optically active trions in both singlet (two electrons from the same valley) and triplet (electrons from different valleys) configurations is favoured in the so-called "darkish" monolayers (WSe$_2$ and WS$_2$ ), i.e when the ground exciton is optically inactive (dark). In contrast, molybdenum ML is characterized by an optically active (bright) ground-state singlet exciton [17, 41, 67-72]. The dielectric environment also plays an important role [67]. Indeed, the existence of a fine structure strongly depends on whether the ML is encapsulated, supported, or suspended. Recently, by comparing the PL spectra measured on the WS2 monolayer deposited on SiO$_2$/Si substrate and on that encapsulated in hBN akes (van der Waals heterostructure), Vaclavkova *et al*. [67] show that the doublet structure (triplet and singlet trions) is not resolved in supported WS2 monolayer (deposited on SiO2/Si substrate ) while it becomes apparent in the van der Waals heterostructure. In the same context, Wang *et al*.[15] show that under circularly polarized excitation, the PL spectra of WSe2 monolayer deposited on a SiO2 substrate is characterized by neutral exciton peak X with full width at half maximum about 10 meV, negative-charged exciton peak ($X_-$ ) appearing 30 meV below the free bright exciton peak(PL FWHM 15 meV) and without any resolved fine structure. Finally, it has also been shown that the triplet trion quickly disappears from the PL spectra with increasing temperature and only the singlet trion peak is observed in the temperature range above 60 K [67]. Here, we cover a regime of high temperature, low doping density, and for deposited TMDCs ML, and consider only the initial singlet bound trion state. Therefore, for the ground (bound) states, the relative trion wave function can be expanded using an auxiliary basis :

$$\zeta_{X_-}(\rho_1,\rho_2) = \sum_{\tilde{n},\tilde{l}} D(\tilde{n},\tilde{l}) \frac{1}{\sqrt{2}} \{\phi_{\widetilde{1s}}(\rho_1)\phi_{\widetilde{nl}}(\rho_2) + \phi_{\widetilde{1s}}(\rho_2)\phi_{\widetilde{nl}}(\rho_1)\}. \quad (A.3)$$

Where, $\phi_{\widetilde{nl}}(\rho,\theta) = \sum_{n,l} C(n,l) \varphi_{n,l}(\rho,\theta)$ are the eigenvalue solution of the exciton hamiltonien $H_X$, $\phi_{\widetilde{nl}}(\rho,\theta)$ is expanded in terms of 2D-hydrogenic state $\varphi_{n,l}(\rho,\theta)$. The number $\tilde{n}$, $\tilde{l}$ refers to the dominant contribution of the coefficients $C(n,l)$ to the excitonic function. Notably, having obtained the trion eigenstates by numerical diagonalization, we can finally calculate the trion energies $E_{BX_-}$, which is conventionally defined as the difference between the PL peaks energies of the neutral and charged excitons. Let us now consider the final trion state. Since the interaction with phonons conserve the particles' spin, only the singlet final (dissociated) trion states will also be discussed. For the unbound electron-exciton X (dissociated trion), the simplest approximation consists in neglecting the interaction between the exciton and the electron. According to this assumption, the unbound electron-exciton wave function and energy have, respectively the following form :

$$\Psi_{X-e} = \frac{1}{\sqrt{2A^2}} \{e^{iq_e\rho_{e_1}} e^{iK_X R_{X_2}} \phi_{\widetilde{1s}}(\rho_2) + e^{iq_e\rho_{e_2}} e^{-iK_X R_{X_1}} \phi_{\widetilde{1s}}(\rho_1)\} \quad (A.4)$$

$$E_{X-e} = E_X + \frac{\hbar^2 K_X^2}{2M_X} + \frac{\hbar^2 q_e^2}{2m_e} \quad (A.5)$$

Now, using the following transformation:

$$\begin{cases} \rho_{e_1} = R_{X_-} + \frac{M_X}{M_{X_-}}\rho_1 - \frac{m_e}{M_{X_-}}\rho_2 \\ \rho_{e_2} = R_{X_-} - \frac{m_e}{M_{X_-}}\rho_1 + \frac{M_X}{M_{X_-}}\rho_2 \\ R_{X_1} = R_{X_-} + \frac{m_e^2}{M_X M_{X_-}}\rho_1 - \frac{m_e}{M_{X_-}}\rho_2 \\ R_{X_2} = R_{X_-} - \frac{m_e}{M_{X_-}}\rho_1 + \frac{m_e^2}{M_X M_{X_-}}\rho_2 \end{cases} \quad (A.6)$$

the unbound electron-exciton $X$ wave function can be writen as :

$$\Psi_{X-e}(R_{X_-},\rho_1,\rho_2) = \frac{1}{\sqrt{2A}} e^{i(q_e+K_X)R_{X_-}} [\Phi_{X-e}(\rho_1,\rho_2) + \Phi_{X-e}(\rho_2,\rho_1)] \quad (A.7)$$

with

$$\Phi_{X-e}(\rho_1,\rho_2) = \frac{1}{\sqrt{A}} exp\left[\frac{i}{M_{X_-}}(M_X q_e - m_e K_X).\left(\rho_1 - \frac{m_e}{M_X}\rho_2\right)\right] \phi_{\widetilde{1s}}(\rho_2) \quad (A.8)$$

If $m_h \to \infty$, $\Phi_{X-e}(\boldsymbol{\rho_1}, \boldsymbol{\rho_2}) = \frac{1}{\sqrt{A}} e^{i(\boldsymbol{q_e} \cdot \boldsymbol{\rho_1})} \phi_{\widetilde{1s}}(\boldsymbol{\rho_2})$, however in the general case the exponential term in Eq.A8 depend on both $\boldsymbol{\rho_1}$ and $\boldsymbol{\rho_2}$. We can define $\boldsymbol{K_{X_-}^f} = \boldsymbol{q_e} + \boldsymbol{K_X}$. In fact, in our system we can define the following transformation:

$$\begin{cases} \boldsymbol{q_e} = \alpha \boldsymbol{K_{X_-}^f} + \gamma \boldsymbol{K} \\ \boldsymbol{K_X} = \beta \boldsymbol{K_{X_-}^f} - \gamma \boldsymbol{K} \end{cases} \quad (A.9)$$

with $\alpha + \beta = 1$ and $\gamma$ so far arbitrary (see below). We obtain :

$$E_{X-e} = E_X + \frac{\hbar^2}{2}\left[\left(\frac{\alpha^2}{m_e} + \frac{\beta^2}{M_X}\right){\boldsymbol{K_{X_-}^f}}^2 + 2\gamma\left(\frac{\alpha}{m_e} - \frac{\beta}{M_X}\right)\boldsymbol{K_{X_-}^f} \cdot \boldsymbol{K} + \gamma^2\left(\frac{1}{m_e} + \frac{1}{M_X}\right)\boldsymbol{K}^2\right] \quad (A.10)$$

The crossed term vanishes for $\alpha = \frac{m_e}{M_{X_-}}$, and $\beta = \frac{M_X}{M_{X_-}}$. Using Eq.A9 and replacing $\alpha$ and $\beta$ by their values, $\Phi_{X-e}(\boldsymbol{\rho_1}, \boldsymbol{\rho_2})$ can be rewriten as :

$$\Phi_{X-e}(\boldsymbol{\rho_1}, \boldsymbol{\rho_2}) = \frac{1}{\sqrt{A}} exp\left[i.\gamma \boldsymbol{K}\left(\boldsymbol{\rho_1} - \frac{m_e}{M_X}\boldsymbol{\rho_2}\right)\right] \phi_{\widetilde{1s}}(\boldsymbol{\rho_2}) \quad (A.11)$$

with

$$\boldsymbol{\rho_1} - \frac{m_e}{M_X}\boldsymbol{\rho_2} = \frac{M_{X_-}}{M_X}\left(\boldsymbol{\rho_{e_1}} - \boldsymbol{R_{X_-}}\right)$$

Taking $\gamma = -1$ leads to an unit transformation of $(\boldsymbol{q_e}, \boldsymbol{K_X}) \rightsquigarrow (\boldsymbol{K_{X_-}^f}, \boldsymbol{K})$ allowing to replace in the following the summations : $\sum_{q_e}\sum_{K_X} \leftrightarrow \sum_K \sum_{K_{X_-}^f}$. Finally, the energy and wave function of the unbound electron-exciton state are given respectively by:

$$E_{X-e}(\boldsymbol{K_{X_-}^f}, \boldsymbol{K}) = E_X + \frac{\hbar^2 {\boldsymbol{K_{X_-}^f}}^2}{2M_{X_-}} + \frac{\hbar^2 K^2}{2\mu} \quad (A.12)$$

$$\Psi_{X-e}(\boldsymbol{R_{X_-}}, \boldsymbol{\rho_1}, \boldsymbol{\rho_2}) = \frac{1}{\sqrt{A}} e^{i\boldsymbol{K_{X_-}^f} \boldsymbol{R_{X_-}}} \zeta_{X-e}(\boldsymbol{\rho_1}, \boldsymbol{\rho_2}) \quad (A.13)$$

where

$$\zeta_{X-e}(\boldsymbol{\rho_1}, \boldsymbol{\rho_2}) = \frac{1}{\sqrt{2A}}\left[e^{-i\boldsymbol{K}\left(\boldsymbol{\rho_1} - \frac{m_e}{M_X}\boldsymbol{\rho_2}\right)} \phi_{\widetilde{1s}}(\boldsymbol{\rho_2}) + e^{-i\boldsymbol{K}\left(\boldsymbol{\rho_2} - \frac{m_e}{M_X}\boldsymbol{\rho_1}\right)} \phi_{\widetilde{1s}}(\boldsymbol{\rho_1})\right]$$

**Appendix B: The dielectric constants of the surrounding environment**

Choosing between the static or high-frequency dielectric constants of the surrounding environment is a subtle problem when one wishes to calculate the exciton and trion optical properties in monolayers TMDCs. In fact, in our study to calculate the up and down conversion rate as well as the energie, we should use the high frequency (infrared) values for the various dielectric constants, rather than static values [73, 74]. This choice is justified because the polar-phonon energies in TMDCs materials are smaller than the exciton binding energies (several hundred of meV) both in the material itself and in the typical substrates or cap layers, including SiO$_2$ and hexagonal boron nitride. For further details see ref. [73, 74].

**Appendix C: Evolution of up-conversion process**

After determining the eigenfunctions and the eigenvalues of the initial and final states and proving that the trion binding energy of TMDs monolayer (WSe$_2$, WS$_2$, MoS$_2$, MoSe$_2$, MoTe$_2$) match with single optical phonon energy ($E^{'}$ or $A_1^{'}$), we proceed now to explicit the calculation details of the up-conversion rate equation (Eq.1 in the main text).

Table III. TMDCs parameters used in the modelling of the phonon assisted up and down -conversion rate

|  | WSe$_2$ | WS$_2$ | MoSe$_2$ | MoS$_2$ | MoTe$_2$ |
|---|---|---|---|---|---|
| $C_z$ (eV) [55] | 0.276 | 0.14 | 0.502 | 0.334 | 0.819 |
| $D_{HP}^e$ (eV.Å$^{-1}$) [49,54] | 2.3 | 3.1 | 5.2 | 5.8 | - |
| $D_{HP}^h$ (eV.Å$^{-1}$) [49,54] | 3.1 | 2.3 | 4.9 | 4.6 | - |
| $r_{eff}$ (Å) [47,54,55] | 48.7 | 42 | 53.2 | 46.5 | 69.5 |
| $m_e$ [19,40,54,55,76,77] | 0.29 | 0.31 | 0.5 | 0.45 | 0.57 |
| $m_h$ [19,40,54,55,76,77] | 0.36 | 0.42 | 0.6 | 0.54 | 0.64 |
| $\rho_d$ ($g.cm^{-2}$) [77] | 3.1 | 2.36 | 2.01 | 1.56 | - |

1. **The LO phonon coupling : Frohlich interaction**

The LO mode corresponds to in-plane longitudinal displacements with the molybdenum (tungsten) atoms moving in phase opposition to sulfur or selenide atoms. The origin of the polar-optical coupling is the polarization density generated by the atomic displacement pattern associated with a LO phonon of in-plane momentum Q. This coupling is governed by the the long-range Frohlich interaction, which is fundamentally affected by the dimensionality of the system [17, 40, 49, 54-56, 59, 75]. It has been shown recently that in TMDs material this interaction is very sensitive to the details of the dielectric environment as well as the dielectric properties of the material [40, 54, 55]. Following the definitions given in Refs. [40, 54, 55] the amplitude of the long-range 2D Frohlich interaction is given by (the same for electrons and hole):

$$V_{LO}(Q) = \frac{C_z}{\epsilon_{eff} + r_{eff}|\mathbf{Q}|} \quad (C.1)$$

where $C_z$ is a material-dependent constant (see table .III and refs [40, 54, 55].)

The dielectric screening of the electric field of LO mode deformations is described by the static dielectric function $\xi_I(Q) = \epsilon_{eff} + r_{eff}|\mathbf{Q}|$ [40, 54, 55]. The role of the dielectric environment is captured by $\epsilon_{eff} = \frac{\epsilon_b + \epsilon_t}{2}$ the average dielectric constant of the surrounding material where $\epsilon_b$ ($\epsilon_t$) is the dielectric constant of the bottom (top) dielectric layer. The screening length $r_{eff}$ depends on the dielectric properties of the material as well as on its thickness. Actually this length can be interpreted as an effective thickness marking the crossover between two screening regimes [40, 54, 55]. Table.III presents its value for different materials. Note that the bare Frohlich interaction strengths $C_z$ differs strongly with TMDCs materials and is higher for Mo based TMDCs as compared to Tungsten-based ones (see table III). In this work, we are interested in small momentum **Q** values as they most significantly contribute to the up-conversion process [17, 40, 49, 54-56, 59, 75]. For small Q values, the Frohlich interaction is very sensitive to the details of the dielectric environment and the choice of substrate material. Indeed, the introduction of a dielectric environment through a substrate or full encapsulation will affect strongly the electric potential induced by the LO phonon, which will be reduced in the long wavelength limit by $\epsilon_{eff}$ thus leading to a strong dependence of the up conversion rate on the detail of the dielectric environment [40, 54, 55].

## 2. The HP phonon coupling : Deformation potential

The Homopolar vibrations ($A_1'$) correspond to out-of plane displacements of the chalcogen atoms in phase opposition, while the transition metal atoms are static in the long wavelength mode. The homopolar (HP) mode couples with the carriers through the lattice deformation potential. We write [17, 40, 49, 54, 56, 77]

$$V_{HP}^j(Q) = \sqrt{\frac{\hbar}{2N M \omega_{Q,HP}}} D_{HP}^j \quad (C.2)$$

$\omega_{Q,HP}$ is the HP phonon frequency. Due to the dispersionless nature of long-wavelength optical phonons, we will neglect their weak dependences on **Q** i.e. we will take $\omega_{Q,HP} = \omega_\lambda$, where $\lambda$ = LO,HP [40, 56], and write correspondingly, $V_{HP}^j(Q) \simeq V_{HP}^j$. $D_{HP}^j$ is the constant zero-order optical deformation potential of electrons (j = $e_1$, $e_2$) or holes (j = h), we assume that $D_{HP}^{e_1} = D_{HP}^{e_2} \neq D_{HP}^h$. M is the total atomic mass within the unit cell and N is the number of unit cells, such that, $NM = \rho_d A$, where A is the quantization area in the monolayer plane, and $\rho_d$ is the mass density. Table III gives $D_{HP}^j$ for different TMDCs materials.

### 3. Matrix element calculation.

The matrix element $F_\lambda^{UC}(Q, K, K_{X_-}^f, K_{X_-}^i)$ reads:

$$F_\lambda^{UC}(Q, K, K_{X_-}^f, K_{X_-}^i) = \langle \Psi_F | H_{X_- -ph}^\lambda | \Psi_I \rangle =$$
$$\sqrt{N_{\lambda,Q}} \langle \Psi_{X-e} | V_\lambda^{e_1}(Q) e^{iQ\rho_{e_1}} + V_\lambda^{e_2}(Q) e^{iQ\rho_{e_2}} - V_\lambda^h(Q) e^{iQ\rho_h} | \Psi_{X_-}^i \rangle \quad (C.3)$$

Using the intial and final wave function explicit in Appendix A, and expressing the in-plane position vectors of the two electrons and hole as a function of the center of mass $R_{X_-}$ and relative coordinate $\rho_{e_i}$, in Eq .A. 6 the resulting matrix element can be rewritten as :

$$F_\lambda^{UC}(Q, K, K_{X_-}^f, K_{X_-}^i) = \sqrt{N_{\lambda,Q}} \times \delta_{K_{X_-}^f, (K_{X_-}^i + Q)} \, \mathbb{M}_{fi}(Q, K) \quad (C.4)$$

In the last equation, the Kronecker $\delta$ represent momentum conservation in up conversion process. $\mathbb{M}_{fi}(Q, K)$ is given by :

$$\mathbb{M}_{fi}(Q, K) = V_\lambda^{e_1}(Q) \mathbb{I}_{e_1}(Q, K) + V_\lambda^{e_2}(Q) \mathbb{I}_{e_2}(Q, K) - V_\lambda^h(Q) \mathbb{I}_h(Q, K) \quad (C.5)$$

where,

$$\mathbb{I}_{e_1}(Q, K) = \langle \zeta_{X-e}(\rho_1, \rho_2) | e^{i\beta_{X_-} Q\rho_1} e^{-iQ\gamma_{X_-}\rho_2} | \zeta_{X_-}(\rho_1, \rho_2) \rangle =$$
$$\sum_{\widetilde{nl}} \frac{D(\widetilde{n,l})}{2\sqrt{A}} \left[ \mathbb{G}_{e_1}(Q, K, \widetilde{1s}, \widetilde{nl}) + \mathbb{H}_{e_1}(Q, K, \widetilde{1s}, \widetilde{nl}) \right] \quad (C.6)$$

by replacing $\zeta_{X-e}(\rho_1, \rho_2)$ and $\zeta_{X_-}(\rho_1, \rho_2)$ with their expression in Eq A.14 and A.3 respectively, $\mathbb{G}_{e_1}(Q, K, \widetilde{1s}, \widetilde{nl})$ and $\mathbb{H}_{e_1}(Q, K, \widetilde{1s}, \widetilde{nl})$ are given by :

$\mathbb{G}_{e_1}(Q, K, \widetilde{1s}, \widetilde{nl}) =$
$\{\langle e^{-iK\rho_1} | e^{i\beta_{X_-} Q\rho_1} | \phi_{\widetilde{1s}}(\rho_1) \rangle \langle e^{i\beta_X K\rho_2} \phi_{\widetilde{1s}}(\rho_2) | e^{-i\gamma_{X_-} Q\rho_2} | \phi_{\widetilde{nl}}(\rho_2) \rangle +$
$\langle e^{-iK\rho_1} | e^{i\beta_{X_-} Q\rho_1} | \phi_{\widetilde{nl}}(\rho_1) \rangle \langle e^{i\beta_X K\rho_2} \phi_{\widetilde{1s}}(\rho_2) | e^{-i\gamma_{X_-} Q\rho_2} | \phi_{\widetilde{1s}}(\rho_2) \rangle\} = \mathbb{M}(Q, K, \widetilde{1s}) \mathbb{N}(Q, K, \widetilde{1s}, \widetilde{nl}) +$
$\mathbb{M}(Q, K, \widetilde{nl}) \mathbb{N}(Q, K, \widetilde{1s}, \widetilde{1s}) \quad (C.7)$
And

$\mathbb{H}_{e_1}(Q, K, \widetilde{1s}, \widetilde{nl}) =$
$\{\langle e^{-iK\rho_2} | e^{-i\gamma_{X_-} Q\rho_2} | \phi_{\widetilde{nl}}(\rho_2) \rangle \langle e^{i\beta_X K\rho_1} \phi_{\widetilde{1s}}(\rho_2) | e^{i\beta_{X_-} Q\rho_1} | \phi_{\widetilde{1s}}(\rho_1) \rangle +$
$\langle e^{-iK\rho_2} | e^{-i\gamma_{X_-} Q\rho_2} | \phi_{\widetilde{1s}}(\rho_2) \rangle \langle e^{i\beta_X K\rho_1} \phi_{\widetilde{1s}}(\rho_2) | e^{i\beta_{X_-} Q\rho_1} | \phi_{\widetilde{nl}}(\rho_1) \rangle\} =$
$\mathbb{D}(Q, K, \widetilde{nl}) \, \mathbb{S}(Q, K, \widetilde{1s}, \widetilde{1s}) + \mathbb{D}(Q, K, \widetilde{1s}) \, \mathbb{S}(Q, K, \widetilde{1s}, \widetilde{nl}) \quad (C.8)$

Where $\beta_{X_-} = \frac{M_X}{M_{X_-}}, \gamma_{X_-} = \frac{m_e}{M_{X_-}}, \beta_X = \frac{m_e}{M_X}$

The matrix elements $\mathbb{N}(Q, K, \widetilde{nl}, \widetilde{n_1\,l_1})$ and $\mathbb{S}(Q, K, \widetilde{nl}, \widetilde{n_1\,l_1})$ are given by:

$$\mathbb{N}(Q, K, \widetilde{nl}, \widetilde{n_1\,l_1}) = \langle e^{i\beta_X K \rho_i} \phi_{\widetilde{nl}}(\rho_i) | e^{-i\gamma_{X\_} Q\rho_i} | \phi_{\widetilde{n_1\,l_1}}(\rho_i) \rangle =$$
$$2\pi \sum_{nl} C(n,l) \sum_{n_1 l_1} C(n_1, l_1) i^{l_1-l} \int_0^\infty \rho_i \varphi_n(\rho_i)\, \varphi_{n_1}(\rho_i) J_{l_1-l}(|\beta_X K + \gamma_{X\_} Q| \rho_i) d\rho_i \quad (C.9)$$

And

$$\mathbb{S}(Q, K, \widetilde{nl}, \widetilde{n_1\,l_1}) = \langle e^{i\beta_X K \rho_1} \phi_{\widetilde{nl}}(\rho_i) | e^{i\beta_{X\_} Q \rho_i} | \phi_{\widetilde{n_1\,l_1}}(\rho_i) \rangle =$$
$$2\pi \sum_{nl} C(n,l) \sum_{n_1 l_1} C(n_1, l_1) i^{l_1-l} \int_0^\infty \rho_i \varphi_n(\rho_i)\, \varphi_{n_1}(\rho_i) J_{l_1-l}(|\beta_{X\_} Q - \beta_X K| \rho_i) d\rho_i \quad (C.10)$$

$J_{l_1-l}(X)$ is the Bessel function of the first kind $\mathbb{M}(Q, K, \widetilde{nl})$, $\mathbb{D}(Q, K, \widetilde{nl})$ are given by:

$$\mathbb{M}(Q, K, \widetilde{nl}) = \langle e^{-iK\rho_i} | e^{i\beta_{X\_} Q\rho_i} | \phi_{\widetilde{nl}}(\rho_i) \rangle = 2\pi\, C(n,l) i^l \int_0^\infty \int_0^\infty \rho_i \varphi_n(\rho_i) J_l(|\beta_{X\_} Q + K| \rho_i)\, d\rho_i \quad (C.11)$$

And

$$\mathbb{D}(Q, K, \widetilde{nl}) = \langle e^{-iK\rho_i} | e^{-i\gamma_{X\_} Q\rho_i} | \phi_{\widetilde{nl}}(\rho_i) \rangle = 2\pi\, C(n,l) i^l \int_0^\infty \int_0^\infty \rho_i \varphi_n(\rho_i) J_l(|K - \gamma_{X\_} Q| \rho_i)\, d\rho_i \quad (C.12)$$

Note that the multidimensional matrix elements are in the end reduced to expressions involving only one dimensional integrals. This is due to the particular form of our trial wavefunctions, for both the trion and exciton electron states, and represents an important numerical improvement.

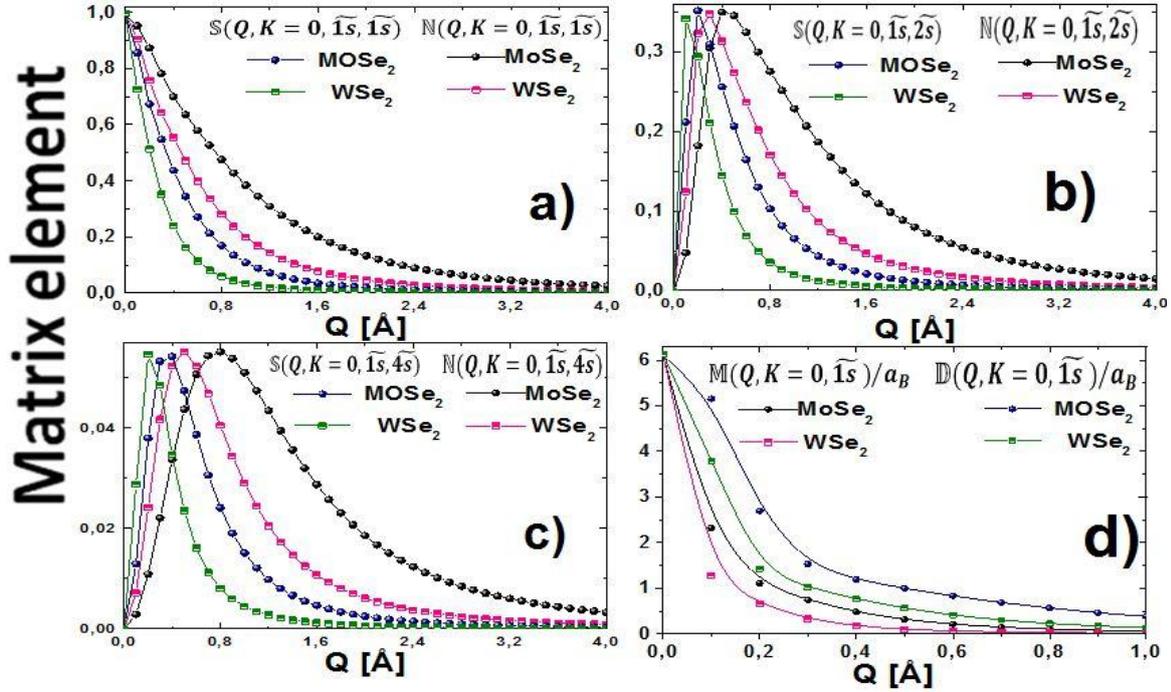

FIG. 5. a-c) The matrix elements $\mathbb{N}(Q, K = 0, \widetilde{1s}, \widetilde{nl})$ and $\mathbb{S}(Q, K = 0, \widetilde{1s}, \widetilde{nl})$ for WSe2 and MoSe2, d) The matrix elements $\mathbb{M}(Q, K = 0, \widetilde{nl})$ and $\mathbb{D}(Q, K = 0, \widetilde{nl})$ for WSe2 and MoSe2.

In order to give an insight on these integrals, we present in Fig. 5 a few of them. That will also allow us to show in a few examples how different material parameters may affect their values. In Fig. 5 (a-d) we plot the matrix element $\mathbb{N}(Q, K = 0, \widetilde{1s}, \widetilde{nl})$ and $\mathbb{S}(Q, K = 0, \widetilde{1s}, \widetilde{nl})$ for $\widetilde{nl} = \widetilde{1s},; \widetilde{2s},, \widetilde{4s}$, as well as $\mathbb{M}(Q, K = 0, \widetilde{1s})$ and $\mathbb{D}(Q, K = 0, \widetilde{1s})$ for $MoSe_2$ and $WSe_2$ monolayer. We see in these figures that the functional dependencies of a given matrix element are roughly the same for either material, with slight variations that can be traced back to the differences in material parameters (see table SIII). We can notice also that the matrix elements $\mathbb{N}(Q, K = 0, \widetilde{1s}, \widetilde{1s})$ ($\mathbb{S}(Q, K = 0, \widetilde{1s}, \widetilde{1s})$) and $\mathbb{M}(Q, K = 0, \widetilde{1s})$ ($\mathbb{D}(Q, K = 0, \widetilde{1s})$) have the most important contribution in the up-conversion process as compared to other transition such as $\widetilde{1s} \to \widetilde{2s}$, $\widetilde{1s} \to \widetilde{3s}$ and $\widetilde{1s} \to \widetilde{4s}$. The matrix element relative to the second electron of the trion $\mathbb{I}_{e_2}(Q, K)$ is obtained just by replacing $\rho_1 \leftrightarrow \rho_2$ in the exponential term of the Eq C.6, while the one describing the interaction of hole with polar optical phonons is given by :

$$\mathbb{I}_h(Q, K) = \langle \zeta_{X-e}(\rho_1, \rho_2) | e^{-i\gamma_{X_-} Q\rho_1} e^{-iQ\gamma_{X_-} \rho_2} | \zeta_{X_-}(\rho_1, \rho_2) \rangle = \sum_{\widetilde{n}\widetilde{l}} \frac{D(\widetilde{n},\widetilde{l})}{2\sqrt{A}} [\mathbb{G}_h(Q, K, \widetilde{1s}, \widetilde{nl}) + \mathbb{H}_h(Q, K, \widetilde{1s}, \widetilde{nl})]$$
(C.13)

Where

$$\mathbb{G}_h(Q, K, \widetilde{1s}, \widetilde{nl}) = \mathbb{H}_h(Q, K, \widetilde{1s}, \widetilde{nl}) = \mathbb{D}(Q, K, \widetilde{nl}) \mathbb{N}(Q, K, \widetilde{1s}, \widetilde{1s}) + \mathbb{D}(Q, K, \widetilde{1s}) \mathbb{N}(Q, K, \widetilde{1s}, \widetilde{nl}) \quad (C.14)$$

$\mathbb{N}(Q, K, \widetilde{1s}, \widetilde{nl})$ and $\mathbb{D}(Q, K, \widetilde{nl})$ are given respectively by Eq C9 and C12.

**Appendix D: Temperature and electron dependence of the down-conversion process**

In Fig.6 (a,b,c,d), we show the temperature and electron density dependences of the down conversion rate for MoSe2 ,WSe2 and MoTe2 due to HP and LO phonon at different electron density $N_e$Ne. The results show that for a given $N_e$, the down conversion rate decreases with increasing temperature, due to the exponential decaying occupation of electronic states with T . Notably, for small for moderate electron density , the electron occupation number$f_e(q_e)$ is only significant when $\frac{\hbar^2 q_e^2}{2m_e} < K_B T$ i.e $q_e = \frac{\sqrt{2K_B T m_e}}{\hbar}$.

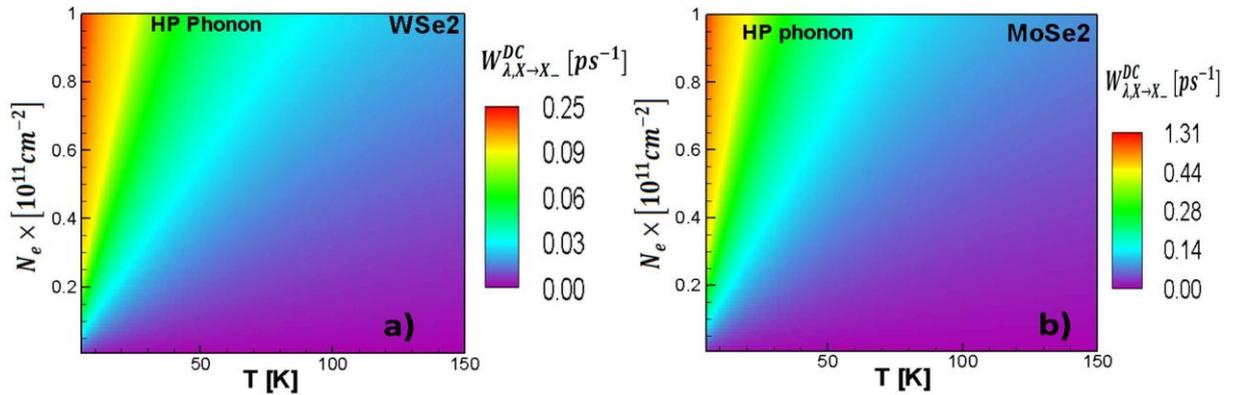

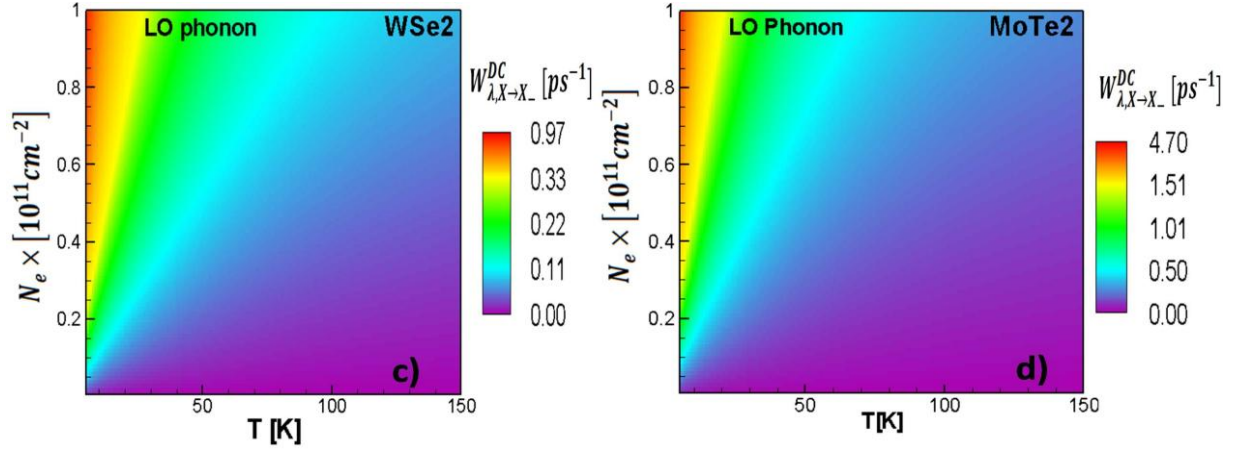

FIG. 6. The down-conversion rate at $K_X = 0$ as a function of the temperature T and electron density $N_e$, for different materials, due to HP phonon a)WSe2 b) MoSe2, and LO phonon c)WSe2 ,d) MoTe2

Note that in this study, we restrict ourselves to the case of low doping density. In fact, for high doping density $N_e > 10^{12} cm^{-2}$ many body effects can influence the exciton (trion) to trion (exciton) down (up) conversion process, such as trion fine structure, excitons (shortwave) plasmon coupling. A detailed study of the role of these many-body effects on the phonon-assisted conversion processes is beyond the scope of this work.

**Appendix E: Exciton radiative lifetime at low and high Temperature.**

The radiative lifetime of the exciton is due to coupling with a continuum of photon states. We consider free 2D Wannier excitons and light propagating perpendicular to the 2D layer, such that in-plane wave vector is conserved during the photon emission [61, 78, 79]. Following the approach of Robert *et al.*[61] and Glazov *et al.*,[78] the radiative decay rate of free exciton is given by:

$$\Gamma^r_{0,X} = \frac{1}{2\tau^r_{0,X}} = \frac{2\pi k_0}{\hbar}\left[\frac{e\hbar v}{n_0 E_X}\right]^2 |\phi_{\widetilde{1s}}(\rho = 0)|^2 \quad (E.1)$$

where $k_0 = \frac{n_0 E_X}{\hbar C}$ is the light wave-vector in the sample, $n_0 = \sqrt{\epsilon_{eff}}$ is the effective optical refraction index of the crystal environment, $v$ is the Kane velocity for TMDCs estimated from a two band models, c is the speed of light. The radiative exciton recombination time increases linearly with the temperature and it is given by:

$$\tau^r_X = \frac{3}{2}\frac{K_B T}{E_0} \tau^r_{0,X} \quad (E.2)$$

where $E_0 = \frac{(\hbar K_0)^2}{2M_X}$ is the kinetic energy of the excitons which decay radiatively

**References**

[1] C. Janisch, Y. Wang, D. Ma, N. Mehta, A. L. Elias, N. Perea-Lopez, M. Terrones, V. Crespi, and Z. Liu, Sci. Rep. **4**, 5530 (2014).

[2] J. Huang, T. B. Hoang, and M. H. Mikkelsen, Sci. Rep. **6**, 22414 (2016).

[3] A. T. Hanbicki, G. Kioseoglou, M. Currie, C. Stephen Hellberg, K. M. McCreary, A. L. Friedman, and B. T. Jonker, Sci. Rep. **6**, 18885 (2016).

[4] K. F. Mak, C. Lee, J. Hone, J. Shan, and T. F. Heinz, Phys. Rev. Lett. 105, 136805 (2010).
[5] L. Cao, MRS BULLETIN **40**, 592 (2015).

[6] P.Tonndorf, R. Schmidt, P. B• ottger, X. Zhang, J. Borner, A. Liebig, M. Albrecht, C. Kloc, O. Gordan, D. R. T. Zahn, S. M. de Vasconcellos, and R. Bratschitsch, Optics Express.**21**,4908 (2013).

[7] D. Xiao, G.B. Liu, W.Feng, X. Xu, and W. Yao, Phys. Rev. Lett.**108**, 196802 (2012).

[8] A. T. Hanbicki, M. Currie, G. Kioseoglou, A. L. Friedman, and B. T. Jonker,Solid State Communications.**203**,16 (2015).

[9] B. Zhu, X. Chen , X. Cui, Sci. Rep. **5**, 9218 (2015).

[10] K. He, N. Kumar, L. Zhao, Z. Wang, K.F. Mak, H. Zhao, and J. Shan,Phys. Rev. Lett.113, 026803 (2014).

[11] D. K. Zhang, D. W. Kidd, and K. Varga , Nano Lett. **15**, 7002 (2015).

[12] A. V. Stier, N. P. Wilson, G. Clark, X. Xu, and S. A. Crooker , Nano Lett **16**, 7054 (2016).

[13] A. Raja, A. Chaves, J. Yu, G. Arefe, H. M. Hill, A. F. Rigosi, T. C. Berkelbach, P. Nagler, C. Schuller, T. Korn, C. Nuckolls, J.Hone, L.E. Brus, T. F. Heinz, D.R. Reichman and A. Chernikov, Nat Commun **8,** 15251 (2017).

[14] A. Chernikov, T. C. Berkelbach, H. M. Hill, A. Rigosi, Y. Li, O. B. Aslan, D. R. Reichman, M. S. Hybertsen, and T. F. Heinz, Phys. Rev. Lett **113**, 076802 (2014).

[15] G. Wang, L. Bouet, D. Lagarde, M. Vidal, A. Balocchi, T. Amand, X. Marie, and B. Urbaszek ,Phys. Rev. B **90**, 075413 (2014).

[16] R. L. Greene, K. K. Bajaj, and D. E. Phelps,Phys. Rev. B **29**, 1807 (1984).

[17] A.M. Jones, H. Yu, J. R. Schaibley, J.Yan, D.G. Mandrus, T.Taniguchi, K.Watanabe , H. Dery,W.Yao and X. Xu, Nature Physics **12**, 323 (2016).

[18] M. Szyniszewski, E. Mostaani, N. D. Drummond, and V. I. Falko,Phys. Rev. B **95**, 081301(R) (2017).

[19] B. Han, C. Robert, E. Courtade, M. Manca, S. Shree,T. Amand, P. Renucci, T. Taniguchi, K. Watanabe, X. Marie, L. E. Golub, M. M. Glazov, and B. Urbaszek, Phys. Rev. X **8**, 031073 (2018).

[20] G. Wang, M. M. Glazov, C. Robert,T. Amand, X. Marie, and B. Urbaszek , Phys. Rev. Lett **115**, 117401 (2015).

[21] M. Manca, M.M. Glazov, C. Robert, F. Cadiz, T. Taniguchi, K. Watanabe, E. Courtade, T. Amand, P. Renucci, X. Marie, G. Wang , B. Urbaszek, Nat. Communications **8**, 14927 (2017).

[22] J. Dong, W. Gao, Q. Han, Y. Wang, J.Qi, X. Yan, M. Sun, Reviews in Physics 4 100026 (2019).

[23] G. Moody, C. K.Dass, K. Hao, C.H. Chen, L.J. Li, A. Singh,K. Tran, G. Clark, X. Xu, G. Berghauser, E. Malic, A. Knorr and X. Li,Nature Communications **6**, 8315 (2015).

[24] J. Jadczak, L. Bryja, J. Kutrowska-Girzycka , P. Kapuscinski , M. Bieniek, Y.-S. Huang and P. Hawrylak, Nat Commun.**10**,107 (2019).

[25] C. Robert, M.A. Semina, F. Cadiz, M. Manca, E. Courtade, T. Taniguchi, K. Watanabe, H. Cai, S. Tongay, B. Lassagne, P. Renucci, T. Amand, X. Marie1, M.M. Glazov, B. Urbaszek, Phys. Rev.MATERIALS.**2**, 011001(R) (2018).

[26] K. Hao, L. Xu, P. Nagler, A. Singh,K. Tran, C. K. Dass, C. Schuller, T. Korn, X. Li, and G. Moody, Nano Lett. **16**, 5113(2016).

[27] E.Downing, L. Hesselink, J. Ralston,R. Macfarlane, Science **273**, 1185 (1996).

[28] T. Fujino,T. Fujima, and T. Tahara,Phys Chem B **109**,15327 (2005).

[29] A. V. Kachynski, A. N. Kuzmin, H. E. Pudavar, and P. N. Prasad,Appl Phys Lett **87**, 023901 (2005).

[30] R. M. Macfarlane, F. Tong, A. J. Silversmith, and W. Lenth ,Appl Phys Lett **52**, 1300 (1998).

[31] X. Luo, M. D. Eisaman, and T. R. Gosnell, Opt Lett **23**, 639 (1998).

[32] F. Auzel, Phys. Rev. B **13**, 2809 (1976).

[33] R. I. Epstein, M. I. Buchwald, B. C. Edwards, T. R. Gosnell, and C. E. Mungan, Nature **377**, 500 (1995).

[34] F. Wang, G. Dukovic, L.E. Brus, T. F. Heinz. Science **308**, 838 (2005).

[35] W. Seidel, A. Titkov, J. P. Andre, P. Voisin, and M. Voos, Phys. Rev. Lett. **73**, 2356 (1994).

[36] R. Hellmann, A. Euteneuer, S. G. Hense, J. Feldmann, P. Thomas, E. O. G• obel, D. R. Yakovlev, A. Waag, and G.Landwehr, Phys. Rev. B **51**, 18053 (1995).

[37] E. Poles, D. C. Selmarten, O. I. Micic,and A. J. Nozik, Appl. Phys. Lett.**75** , 971 (1999).

[38] P. P. Paskov, P. O. Holtz, B. Monemar, J.M. Garcia, W. V. Schoenfeld, and P. M. Petro , Appl. Phys. Lett. **77**, 812 (2000).

[39] S. L. Chen, J. Stehr, N. Koteeswara Reddy, C.W. Tu, W. M. Chen, and I. A. Buyanova, , Appl. Phys. B 108, 919 (2012).

[40] D. Van Tuan, A. M. Jones, M. Yang, X. Xu, and H. Dery, Phys.Rev.lett. **31**, 122 (2019)

[41] J. Jadczak, J. Kutrowska-Girzycka, P. Kapuscinski, Y. S. Huang, A .W_ojs and L .Bryja, Nanotechnology **28**, 395702 (2017).

[42] K.F. Mak, K. He, C. Lee, G. H. Lee, J. Hone, T. F.Heinz, and J. Shan, Nature Materials , **12**, 211 (2013).

[43] J. W. Christopher, B. B. Goldberg, and A. K. Swan, Scientific Reports **7**, 14062 (2017).

[44] G.Plechinger, P. Nagler, J. Kraus, N. Paradiso, C. Strunk, C. Schuller,and T. Korn, Phys. Status Solidi RRL **9**, 457 (2015).

[45] The TMD monolayer has $D_{3h}$ point symmetry and $P\bar{6}m2$ (187 or $D_{3h}^1$) space group. With three atoms in the unit cell, single-TMDC layer has nine phonon branches. In fact of the nine phonon modes in MLTMDs, six belong to the optical branches ($E^{'}$ (LO,TO), $E^{''}$ (LO,TO), $A_1^{'}$, $A_2^{''}$ . Two of which are strongly coupled to spin-conserving scattering of electrons or holes . The longitudinal optical (LO) and Homopolar phonons. We neglect the transverse optical and l modes due to their weak coupling at the Γ point, $Q \to 0$ .

[46] A. Hichri, I. Ben Amara, S. Ayari, and S. Jaziri, J. Appl. Phys. **121**, 235702 (2017).

[47] T. C. Berkelbach, M. S. Hybertsen, and D. R. Reichman, Phys. Rev. B **88**, 045318 (2013).

[48] R. Y. Kezerashvili, S. M. Tsiklauri ,Few-Body Systems **58**,18 (2016).

[49] Z. Jin, X. Li, J. T. Mullen, and K. W. Kim, Phys. Rev. B, **90**, 045422, (2014).

[50] J. Jadczak, A. Delgado, L. Bryja,Y. S. Huang, and P. Hawrylak, Phys. Rev. B **95**, 195427 (2017).

[51] H. Dery and Y. Song, Phys. Rev. B **92**, 125431 (2015).

[52] C. Robert, R. Picard, D. Lagarde, G. Wang, J. P. Echeverry, F. Cadiz, P. Renucci, A. Hogele, T. Amand, X. Marie,I. C. Gerber, and B. Urbaszek, Phys. Rev. B **94**, 155425 (2016).

[53] I. G.Lezama, A.Arora, A. Ubaldini, C.Barreteau, E. Giannini, M. Potemski and A.F. Morpurgo, Nano Lett. **15**, 2342 (2015).

[54] M. Danovich, I. L. Aleiner, N. D. Drummond, V. I. Falko, IEEE Journal of Selected Topics in Quantum Electronics **23**,1 (2016) .

[55] T. Sohier, M. Calandra, and F. Mauri, Phys. Rev. B **94**, 085415 (2016).

[56] K. Kaasbjerg, K. S. Thygesen, and K. W. Jacobsen,Phys. Rev. B **85**, 115317 (2012).

[57] A. Molina-Sanchez and L. Wirtz, Phys. Rev. B **84**, 155413 (2011).

[58] H. Sahin, S. Tongay,S. Horzum,W. Fan, J. Zhou, J. Li, J. Wu, and F. M. Peeters, Phys. Rev. B **87,** 165409 (2013).

[59] A. Thilagam, J. Appl. Phys. **120**, 124306 (2016).

[60] B.Miller, J. Lindlau, M.Bommert, A. Neumann, H. Yamaguchi,A. Holleitner, A. H• ogele and U. Wurstbauer, Nature Communications **10**, 807 (2019).

[61] C. Robert, D. Lagarde, F. Cadiz, G. Wang, B. Lassagne, T. Amand, A. Balocchi, P. Renucci, S. Tongay, B. Urbaszek and X. Marie,Phys. Rev. B **93**, 205423 (2016).

[62] T. Godde, D. Schmidt, J. Schmutzler, M. Abmann, J. Debus, F. Withers, E. M. Alexeev, O. Del Pozo Zamudio, O. V. Skrypka, K. S. Novoselov, M. Bayer, and A. I. Tartakovskii, Phys. Rev. B **94**, 165301 (2016).

[63] A. Hichri, I. Ben Amara, S. Ayari, and S. Jaziri, J. Phys. Condens. Matt **29**, 435305 (2017).

[64] L. V. Keldysh, Pisma Zh. Eksp. Teor. Fiz. JETP Lett.**29**, 658 (1979).

[65] P. Cudazzo, I. V. Tokatly, and A. Rubio, Phys. Rev. B **84**, 085406 (2011).

[66] D. Andronikov, V. Kochereshko, and A. Platonov,T. Barrick and S. A. Crooker,G. Karczewski,Science **214** ,832 (2000).

[67] D. Vaclavkova, J. Wyzula, K . Nogajewski, M. Bartos, A. O. Slobodeniuk, C. Faugeras, M. Potemski, M .R Molas, Nanotechnology **29**, 325705 (2018).

[68] G. Plechinger , P. Nagler , A. Arora, R.Schmidt, A.Chernikov , A. Granados del Aguila, P.C.M. Christianen, R. Bratschitsch, C. Schuller and T. Korn, Nature Communications **7**, 12715 (2016).

 [69] M. Koperski, M. R. Molas, A. Arora, K. Nogajewski, A. O. Slobodeniuk, C. Faugeras and M.Potemski, Nanophotonics **6**, 1289 (2017).

[70] A. Singh, K. Tran, M. Kolarczik, J. Seifert, Y. Wang, K. Hao, D. Pleskot, N. M. Gabor, S. Helmrich, N. Owschimikow, U. Woggon, and X. Li, Phys. Rev. Lett. **117**, 257402 (2016).

[71] E. Courtade, M. Semina, M. Manca, M. M. Glazov, C. Robert, F. Cadiz, G. Wang, T. Taniguchi, K. Watanabe, M. Pierre, W. Escoffier, E. L. Ivchenko, P. Renucci, X. Marie, T. Amand, and B. Urbaszek , Phys. Rev. B **96**, 085302 (2017).

[72] M.R. Molas, K. Nogajewski, A. O. Slobodeniuk, J. Binder, M. Bartos,a and M. Potemski, Nanoscale **9**, 13128 (2017).

[73] D. V. Tuan, M.Yang, and H. Dery,Phys. Rev. B **98**, 125308 (2018).

[74] M. M. Glazov and A.Chernikov,Physica status solidi b **255**,12 (2018).

[75] X. Li, J. T. Mullen, Z. Jin, K. M. Borysenko, M. Buongiorno Nardelli, and K. W. Kim, Phys. Rev. B **87**, 115418 (2013).

[76] A. Kormanyos , G. Burkard , M. Gmitra , J.Fabian , V. Z´olyomi , N. D Drummond and V. Falko, 2D Materials **2**, 022001 (2015).

[77] A. Thilagam, Journal of Applied Physics **119**, 164306 (2016).

[78] M. M. Glazov, E. L. Ivchenko, G. Wang, T. Amand, X. Marie, B. Urbaszek, B. L. Liu, Phys. Status Solidi B, **252**, 2349 (2015).

[79] A. J. Goodman, A. P. Willard, W. A. Tisdale, Phys. Rev. B **96**, 121404(R) (2017).